\documentclass[%
 reprint,
superscriptaddress,
 amsmath,amssymb,
 aps,
]{revtex4-2}

\usepackage{graphicx}% Include figure files
\usepackage{dcolumn}% Align table columns on decimal point
\usepackage{bm}% bold math
\usepackage{xcolor} % color text
\begin{document}

\preprint{APS/123-QED}

\title{Self-organization and shape change by active polarization in nematic droplets}% Force line breaks with \\

\author{Fabian Jan Schwarzendahl}
\affiliation{Department of Physics, University of California Merced,  Merced, California 95343, USA}
\affiliation{Institut für Theoretische Physik II: Weiche Materie, Heinrich-Heine-Universität Düsseldorf, 40225 Düsseldorf, Germany}
%5200 N. Lake Road: street address usually not needed
 %\altaffiliation[Also at ]{Physics Department,  University of California, Merced}%Lines break automatically or can be forced with \\
 \author{Pierre Ronceray}
 \affiliation{Aix Marseille Univ, Universit\'e de Toulon, CNRS, CPT, Turing Center for Living Systems, Marseille, France}

 \author{Kimberly L. Weirich}
  \affiliation{Department of Materials Science and Engineering, Clemson University, Clemson, SC 29634, USA}
\author{Kinjal Dasbiswas} %\email{Second.Author@institution.edu}
\affiliation{Department of Physics, University of California Merced,  Merced, California 95343, USA}

\date{\today}

\begin{abstract}
Active forces occurring within cells can drive crucial biological processes that involve spontaneous organization and shape change, such as cell division.  Motivated by recent \emph{in vitro} experiments of nematic droplets of cytoskeletal filaments and motors that self-organize and divide, we present a minimal hydrodynamic model that combines the nonequilibrium kinetics of motor-filament interactions with equilibrium nematic phase separation. The motors organize within droplets and structure filaments into polarized aster defects. At large motor activity, they can even deform or divide the droplet, or form multi-aster chains of droplets. Our predicted phase diagram recapitulates these experimentally observed shapes. 
\end{abstract}
\maketitle

\textit{Introduction}. 
Active mechanical forces enable living systems, particularly animal cells, to move, change shape, organize components, and divide. Subcellular cytoskeletal assemblies, comprising polar filaments and molecular motors that transduce biochemical reactions to generate active mechanical forces, drive these processes~\cite{gardel_15}. %\textcolor{red}{cite this here?:\cite{bernheim2018living}, maybe some other review?} KD: I think the review I now cited is a better fit. 
Understanding the general physical principles of living matter provides insight into cell biology, as well as guides the engineering of artificial cells that exhibit spatiotemporal organization of components and spontaneous shape change characteristic of cell division~\cite{Noireaux2011}. Model \emph{in vitro} systems of purified cytoskeletal proteins, which capture elements of cell biological phenomena with only a fraction of the  biochemical complexity occurring \emph{in vivo}, exhibit a rich array of  collective phenomena~\cite{SoareseSilva2011, sanchez2012spontaneous} that motivates bio-inspired active matter theory~\cite{Marchetti2013}.

 Recently, phase segregated macromolecular droplets have emerged as model systems to investigate spatiotemporal organization in biological cells and phase separation has been proposed as a possible primitive means of subcellular organization in protocells~\cite{Hyman2014}.  These macromolecular liquids are typically composed of disordered proteins and nucleic acids, and consequently, internal droplet order as well as motor activity are absent. In contrast, recent experiments indicate that biopolymers with high aspect ratio, such as cytoskeletal filaments, can form phase separated droplets with orientational order, as in liquid crystals, because of the alignment of filaments in the dense phase~\cite{DeGennesText}. This nematic order confers an equilibrium spindle shape to these droplets~\cite{Weirich17,Brugues2014}, known as tactoids, %Structural order, such as filament alignment, generically forms defects in soft materials~\cite{Rudnick95}, which can be reflected in the equilibrium shapes. 
 which arises from a competition of droplet surface tension, the tendency of the filaments to align with the interface, and elasticity arising from bulk nematic order~\cite{Taraskin02, vdS03}. 
 
 The ordered structure in biomolecular fluids influences the emergence of  collective phenomena in active systems~\cite{Marchetti2013}. When confined to a droplet, active forces lead to non-equilibrium phenomena such as droplet shape change~\cite{Giomi2014,Zwicker2017, vutukuri2020,wang2019,li2019,takatori2020,singh2019,ruiz2019}, motility~\cite{Tjhung2012,ziebert2012} and  dynamics governed by the geometry of the confining droplet~\cite{Keber2014}. In addition to these active nematic fluids, the directed ``walking'' of motors on filaments, along with motor-based filament crosslinking, can lead to \emph{polar} order, where filaments prefer to point in the same direction at the mesoscopic scale, and form self-organized defect structures, such as asters and vortices~\cite{Surrey2001}. Such polar active states are fruitfully described by hydrodynamic theories~\cite{Lee2001, Kruse2004, Sankararaman2004,Aranson2006, Ahmadi2005,Ziebert2005, gowrishankar2016nonequilibrium}. In recent experiments, myosin motors were shown to self-organize at the midplane of the aforementioned actin-filament based nematic droplets~\cite{Weirich19}. When sufficiently active myosin motors are present, they deform the droplet, even splitting it into two. A simple free energy-based model of a nematic droplet, considering only the mutual alignment of the filaments and motors and adhesion of the droplet to the motor complex, was invoked to capture these key behaviors, but relied on arguments specific to the shape and structure of both the motor complexes and  the droplet~\cite{Weirich17}. On the other hand, active mechanical forces, specifically the directed sliding of filaments by motors leading to their sorting by polarity~\cite{gardel_15}, are expected to play a role in the dynamics of the organization. The effects of such active emergence of local polar order and associated defects, within a nematic droplet with a preferred orientational axis, have not yet been theoretically explored.

In this Letter, we combine continuum modeling that describes the structure of  equilibrium nematic droplets with an active mechanical model for how motors move and  slide filaments according to polarity, while self-organizing into localized asters. Using numerical simulations, complemented by theoretical analysis, we show that the resulting motor-filament self-organization destabilize droplets, giving rise to a rich array of experimentally observed structures including deformed, divided and multi-lobed droplets that can be generated by tuning one of two motor activity-dependent parameters.

\begin{figure}[b]
    \centering
    \includegraphics[width=\columnwidth]{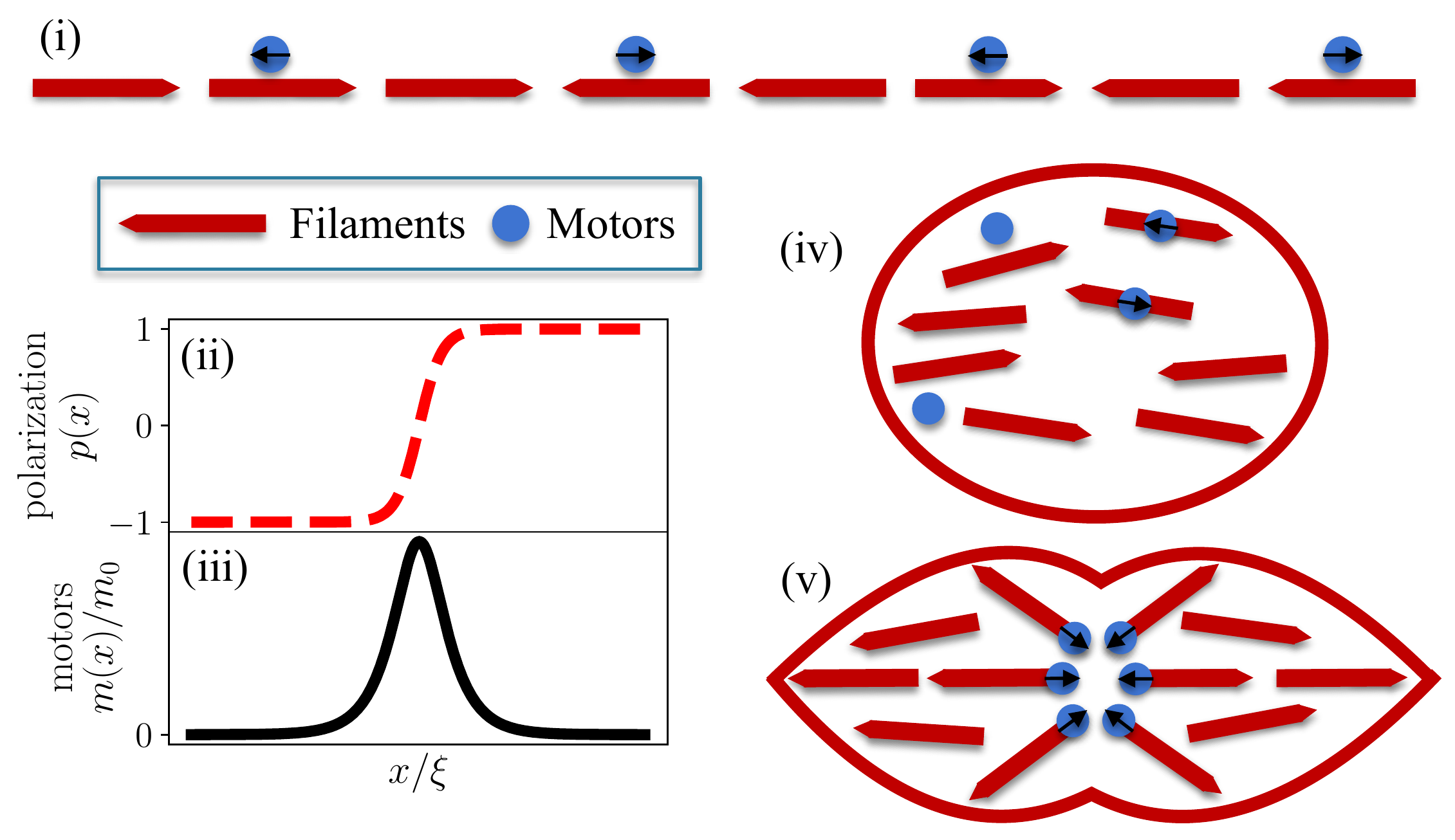}
   \caption{Illustration of steps in the active polarity-sorting model.
   (i) Motors ``walk'' towards barbed ends of filaments which they slide in the opposite direction as a result of momentum conservation.
   (ii)  Polarization and (iii) motor density at steady state as a function of position in a one-dimensional model (Eqs.~\ref{eq:1dmsolution}-\ref{eq:1dpsolution}) that shows motor localization to center of bundle by polarity sorting.
   (iv) In the 2D nematic droplet model, the filaments  are initially oriented along the long axis without any polarity preference. Motors bind to filaments and advect them according to their polarity. (v) At steady state, motors gather at the droplet midplane and sort filaments into an aster that pinches the droplet.  
   }
    \label{fig:sketch1d}
\end{figure}

\textit{1D polarity sorting model.} To build up intuition on the role of active forces in motor self-organization, we first examine a simplified one dimensional setup of motors interacting with filaments, as sketched in Fig.~\ref{fig:sketch1d}(i). Here, red lines depict actin filaments, blue circles show myosin II motors, while the black arrows indicate the motors respective direction of motion. This scenario arises, for instance, in contractile actin bundles \cite{Kruse2003, Stachowiak2012} and when actin filaments are locally oriented along a nematic director \cite{Kumar2018}. Since  motors walk towards a specific end of the polar filament (the barbed or plus end), we consider a density of left and right pointing filaments, denoted by $n_+$ and $n_-$ respectively. Momentum balance for the motor-filament system requires that a right (left) pointing filament is pushed to the right (left) by the motor as the motor moves in a single direction along the polar filament. The active motion of motors of density $m$ thus gives rise to mass fluxes of both motors and filaments. In the dilute regime, the fluxes of right and left pointing filaments can be expressed as $J_+= \zeta n_+ m$ and $J_-= -\zeta n_- m$ respectively, where $\zeta$ is a parameter related to the active motor force. The net flux of motors is written as $- v_{0} (n_+ - n_-) m$, where $v_{0}$ is the self-propulsion velocity of motors.

Including the diffusion of filaments and motors with coefficients $D$ and $D_m$ respectively, the
flux conservation equations can be written in terms of filament density $\rho=n_{+}+n_{-}$ and the 1D polarization $p=(n_{+}-n_{-})/\rho_0$ of filaments, where $\rho_0=\langle \rho \rangle$ is the average density, giving
\begin{align}
\partial_{t} \rho&=D\partial_{x}^{2} \rho - \zeta\rho_0 \partial_{x}(p m) ,
\label{eq:rho1d}
\\ 
\partial_{t} p&= D\partial_{x}^{2} p  - \zeta/\rho_0 \partial_{x}(\rho m),
\label{eq:p1d}
\\
\partial_t m &= D_m\partial_x^2 m + v_0 \partial_x (m p).
\label{eq:motor1d}
\end{align}
The steady-state solution of Eqs.~\eqref{eq:p1d}-\eqref{eq:motor1d}, (see SI I-II), is the one dimensional equivalent of an aster, where the populations of right and left pointing filaments are completely sorted to the right and the left of the aster, and is shown in Fig.~\ref{fig:sketch1d}(ii)-(iii). Note that in this minimal model, we have not considered the usual role of myosin motors as active crosslinkers which create pairwise forces on anti-parallel filaments, leading to self-straining flows proportional to local polarization \cite{Fuerthauer2019}. This simple model reproduces the key experimental observation that motors migrate to the center of the  droplet, and predicts that they induce strong polarity sorting.

\textit{2D nematic droplet with motor-induced polarization}.
To explore this prediction of active polarity sorting in nematic droplets and its implications for droplet shape, we build a more realistic hydrodynamic description for a suspension of filaments and motors %of density, $\psi$, and  $m$, respectively, 
on a frictional substrate that damps out large-scale fluid flows. In contrast with a thin fluid film, the filaments we seek to describe aggregate into droplets with free interfaces that separate the high density nematic from the low density isotropic phases, depicted in Fig.~\ref{fig:sketch1d}(iv). This is conveniently described by a nondimensional ``phase field'' corresponding to filament density, $\psi$, where $\psi > 0$ ( $\psi < 0$) describes the interior (exterior) of the droplet. Filaments in the high density droplet interior align in orientation,  described by the 2D nematic order parameter, $Q_{ij}$. These two ingredients result in nematic droplets with tactoid shape at equilibrium \cite{Ludwig20}. The active motion of filaments and polarity sorting induced by motors results in a net polarization within the droplet.  Unlike in the 1D model,  polar order in 2D is non-conserved and can be induced by motor-driven torques or relaxed by rotational diffusion. Generalizing Eq.~\eqref{eq:rho1d} - \eqref{eq:motor1d} to 2D, and observing usual principles of conservation and symmetry \cite{Marchetti2013}, we obtain the dynamical equations,
\begin{align}
    &\partial_{t} m=D_m \bm{\nabla}^{2} m+v_{0} \bm{\nabla} \cdot (m \bm{p})+ k_{\mathrm{on}} \psi\theta(\psi)-k_{\mathrm{off}} m,
            \label{eq:c_equation}
    \\     
    &\partial_{t} \psi=\Gamma_{\psi} \bm{\nabla}^{2} \frac{\delta F}{\delta \psi} - \zeta \bm{\nabla} \cdot (m \bm{p}),
        \label{eq:psi_equation}
    \\ 
    &\partial_{t} \bm{p}= -\zeta_{0} \bm{\nabla}\psi
    -\zeta_{p} \bm{\nabla} m  
      -\Gamma_{p} \frac{\delta F}{\delta \bm{p}},
        \label{eq:p_equation}
    \\ 
    &\partial_{t} Q_{ij}=-\Gamma_{Q} \frac{\delta F}{\delta Q_{ij}}.
    \label{eq:Q_equation}
\end{align}

 Eq.~\eqref{eq:c_equation}-\eqref{eq:psi_equation} describe the conservation of motors and filaments, respectively, and include both active and passive fluxes on the right side. The motor flux includes diffusion in the first term, and active motion of motors with a propulsion velocity $v_0$ in the second, %is the average  density inside the droplet.}. 
 Additionally, we include binding-unbinding kinetics for the motors (third and fourth term in Eq.~\eqref{eq:c_equation}), where motors bind with rate $k_{\mathrm{on}}$ wherever filaments exist (expressed by $\theta(\psi)$, the Heaviside step function) from a  ``reservoir'' of free motors in solution, and unbind with rate $k_{\mathrm{off}}$, but are not restricted to diffuse within the droplet. Eq.~\eqref{eq:psi_equation} includes a flux created by a free energy of inter-filament interactions (full form given below) as well as active flux induced by motors advecting filaments in their polarity direction.  Eq.~\eqref{eq:p_equation} describes the induction of polarization by torques caused by gradients in motor and filament density, and its relaxation through rotational diffusion. Seen in the framework of the Toner-Tu hydrodynamic theory that describes the flocking of polar active particles~\cite{Toner1995}, where $\bm{p}$ has the status of both an orientational order parameter as well as a fluid velocity, the $-\zeta_{0} \nabla \psi$ gives the gradient of pressure, and $-\zeta_{p} \nabla m$ is the gradient of an active stress created by motors \cite{Husain2017}. Terms similar to this latter also arise in the theory of chemotactic colloids \cite{liebchen2018synthetic} and equilibrium polar liquid crystals \cite{Kung06},in both of which cases a chemical concentration can guide polarization. . Microscopically, the $\zeta_{p}$ term captures the preferred orientation of filaments towards regions with higher motor density by the binding and crosslinking by motors, while $\zeta$ originates from the sliding forces exerted by motors. Note that the $\zeta$ specifies both polarization and active flux in the one dimensional model, whereas the $\zeta_p$ term arises in the two dimensional model because of the additional orientational degree of freedom. The parameters $v_{0}$, $\zeta$ and $\zeta_{p}$ then all depend on motor activity but can in principle be varied independently by tuning motor properties such as  size, shape, processivity and crosslinking.   We assume in Eq. \eqref{eq:Q_equation} that the nematic order is strong and arises from equilibrium forces. The timescales for relaxation towards equilibrium  are specified by the ``mobility'' coefficients, $\Gamma_{\psi}$, $\Gamma_{p}$ and $\Gamma_{Q}$.  

The equilibrium dynamics assume a coupled phase transition in  $\psi$ and nematic order $\bm{Q}$, and a relaxation of $\bm{p}$ included in the total free energy, 
\begin{subequations}
\begin{align}
    F=&F_{\psi}+F_{p}+F_{Q}+F_{\mathrm{int}},
        \label{eq:Full_freeE}
    \\ 
    F_{\psi}=&-\frac{\nu_{2}}{2} \psi^{2}%-\frac{\nu_{3}}{3} \psi^{3}
    +\frac{\nu_4}{4} \psi^{4}+\frac{B}{2}(\bm{\nabla} \psi)^{2},
            \label{eq:psi_freeE}
    \\
    F_{p}=&  \frac{\alpha_p}{2} |\bm{p}|^2
    +\frac{\beta_p}{2} |\bm{p}|^4
    +\frac{\kappa_p}{2}(\bm{\nabla}\cdot \bm{p})^2,
            \label{eq:p_freeE}
    \\
    F_{Q}=&-\frac{\alpha_{Q}}{4} Q_{ij}^{2} +\frac{\beta_{Q}}{16} Q_{ij}^{2} Q_{kh}^{2}+\frac{\kappa_Q}{2}\left(\partial_{k} Q_{i k}\right)^{2},            \label{eq:Q_freeE}
    \\
    F_{\mathrm{int}}=&\frac{C}{4} Q_{i j}^{2} \psi(\psi-1)+A \partial_{i} \psi \partial_{j} Q_{i j}
    +\Omega Q_{i j} p_{i} p_{j}.
        \label{eq:int_freeE}
\end{align}
\end{subequations}
The density free energy Eq.~\eqref{eq:psi_freeE} models phase separating droplets according to standard Cahn-Hilliard dynamics. 
Ignoring corrections for curved interfaces \cite{Everts2016}, we use droplet surface tension (``line tension'' in 2D) as $\gamma= \frac{2\sqrt{2 B \nu_2^3}}{3\nu_4}$ and its interfacial width, $\xi =\sqrt{2 B/\nu_2}$ \cite{safran2018statistical}.
The free energy for the polarization, Eq.~\eqref{eq:p_freeE}, includes two relaxation terms, $\alpha_p, \beta_p > 0$ (corresponding to lack of spontaneous polar order) 
and an elastic term, $\kappa_p$. 
Eq.~\eqref{eq:Q_freeE} is the Landau–de Gennes free energy for the nematic order, $Q_{ij}$, with elasticity $\kappa_Q$. All equilibrium couplings between fields are written in Eq.~\eqref{eq:int_freeE}, where the first term controls the density-driven isotropic--nematic transition and induces nematic order within the droplet.  
The second term  is a ``weak anchoring'' that aligns the nematic parallel to the droplet interface for $A>0$. The third term aligns the polarization with the nematic order. 
\begin{figure}
    \centering
    \includegraphics[width=0.5\textwidth]{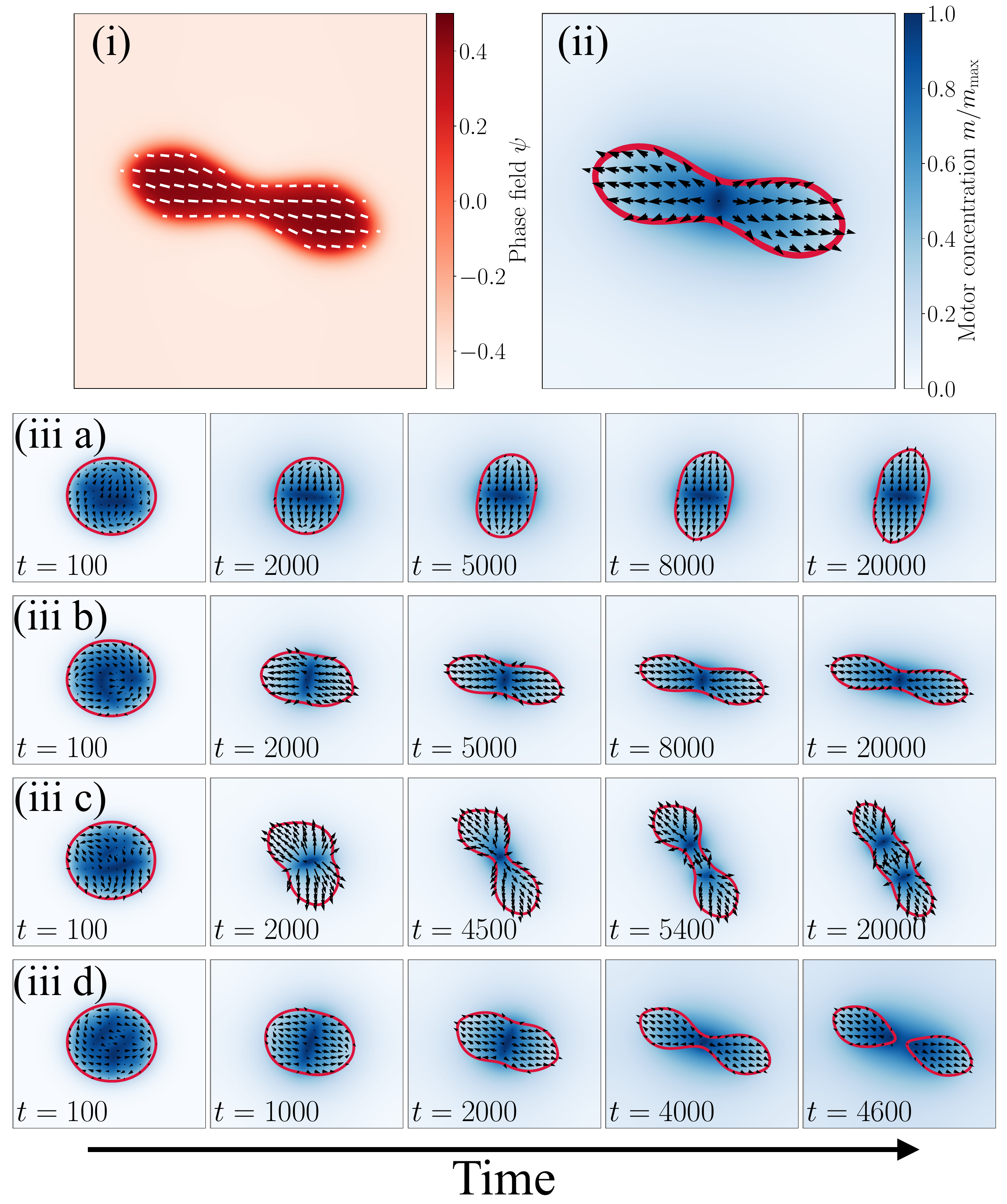}
    \caption{(i) and (ii) A representative numerical steady state solution of our active droplet model. (i) Red gradient indicates the density $\psi$ and the white lines indicate the nematic director field. (ii) Blue gradient, black arrows and the red line indicate the motor concentration,  the  polarization direction and  the droplet boundary respectively.
    (iii): Time series for different droplet types observed in our simulations: (iii a) single tactoid with centering aster; (iii b) two connected tactoids with centering aster; (iii c) three connected tactoids with two asters; (iii d) fully divided droplet. The number in each snapshot shows the simulation time.
    }
    \label{fig:kinetic_pathway}
\end{figure}
Overall, this model recapitulates the key elements of nematic phase separation of the filaments and their coupling with motor activity.

\begin{figure*}
    \centering
    \includegraphics[width=1.0\textwidth]{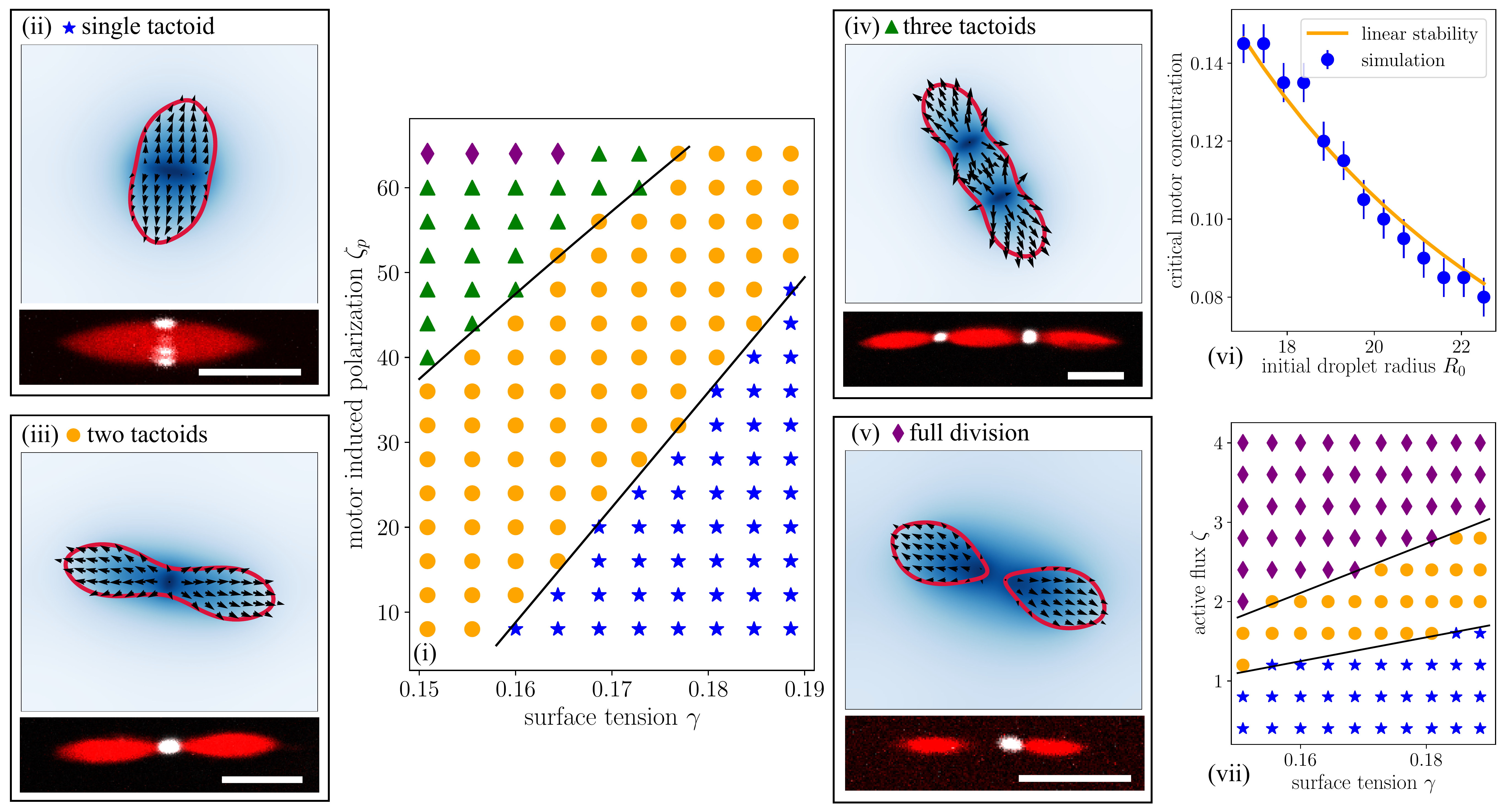}
    \caption{(i) Nonequilibrium phase diagram from our simulations for varying surface tension $\gamma$ and $\zeta_p$ while keeping $\zeta=1.5$ fixed. The solid line is a fit to our theoretical analysis. (ii)-(v) Different steady state shapes observed in our simulations and experiments (scale bars are $5 \mu$m). 
    For the simulation results (ii)-(v) (top) colorcode is the motor concentration $c$ as, black arrows are the polarization, $\bm{p}$, and the red contour shows the shape of the droplet.  We classify the droplet shapes as follows:
    (ii) single tactoid with a centering aster; (iii) two tactoids with one aster; (iv) three connected tactoids with two asters; (v) fully divided droplet. (vi) Critical motor concentration for the transition from one to two asters as a function of initial droplet size. (vii) Nonequilibrium phase diagram showing the different steady states observed in our simulations for varying activity $\zeta$ and surface tension $\gamma$ while keeping $\zeta_p=5.0$ fixed. The solid line is a fit to the prediction from our theoretical analysis. 
    }
    \label{fig:combined_fig}
\end{figure*}

\textit{Results}. We now employ numerical simulations to explore the consequences of motor activity on the dynamics and morphology of phase-separating nematic droplets.
Starting from an initially circular droplet of radius $R_{0}$,  we integrate Eqs.~\ref{eq:c_equation}-\ref{eq:Q_equation} until a non-equilibrium steady-state is reached (see SI III for simulation details and parameters).  Figure~\ref{fig:kinetic_pathway} presents typical simulation results. We  obtain an elongated droplet with the nematic aligned along its long axis, as shown by the density and nematic director plots for a typical case in Fig.~\ref{fig:kinetic_pathway}(i). The motor density accumulates at the core of the aster it induces with an outward polarization as shown in Fig.~\ref{fig:kinetic_pathway}(ii). We first explore the interplay between motor-generated active forces, which tend to distort the equilibrium structure, and surface tension which resists such deformation. To this aim, we perform simulations with varying motor induced polarization, $\zeta_p$, relative to surface tension, $\gamma$.
The resulting time sequences of observed droplet shapes, starting from an unpolarized droplet,  are shown in Fig.~\ref{fig:kinetic_pathway}(iii).  When $\zeta_p$ is low compared to surface tension, Fig.~\ref{fig:kinetic_pathway}(iii a), motors localize towards the center to form an aster, which only slightly polarizes and deforms the droplet. At intermediate $\zeta_p$ (iii b), motors localize more strongly, resulting in a stronger elongation and the appearance of a constriction of the droplet around the midplane between strongly polarized lobes. Increasing  the $\zeta_p$ further can lead to two distinct scenarios: In (iii c), the aster divides and a third, central lobe with strong polarity gradient emerges between two constrictions. In (iii d), finally, $\zeta_p$ is sufficiently strong to induce full division of the initial droplet into two polarized daughter droplets. The short-time dynamics of this model thus reproduces the motor centering, aster formation and polarity sorting features of the 1D model (Fig.~\ref{fig:sketch1d}), while at longer times a complex diversity of morphologies emerges from the interplay between surface tension and motor activity.

To rationalize this rich phenomenology, we build a morphological phase diagram  in Fig.~\ref{fig:combined_fig}(i) in $\zeta_{p}$--$\gamma$ %systematically varying the motor-induced polarization, $\zeta_p$, relative to surface tension,
while keeping other parameters constant. Each distinct steady state structure is shown in
Fig.~\ref{fig:combined_fig}(ii)-(v) (top). Interestingly, each of these morphologies corresponds to experimentally observed shapes, as shown in Fig.~\ref{fig:combined_fig}(ii)-(v) (bottom). 
%Experiments were performed using droplets constituted of actin filaments together with myosin II motors (see SI for details \textcolor{red}{[TODO]}).  
Confirming the qualitative findings described in Fig.~\ref{fig:kinetic_pathway}(iii), at medium to high surface tension and low $\zeta_p$, we find motors form asters in the midplane of the undeformed droplet (blue stars), whereas higher $\zeta_p$ increases the influence of the centered aster on droplet shape, pinching it into  two lobes (orange circles). This is consistent with the experimental observation that motors always localize to droplet midplane, but only deform the droplet when there are more active motors~\cite{Weirich19}. Qualitatively, the motors at the aster core splay the filaments, which are anchored to the interface, and which therefore results in an inward pinching of the interface. 
The motor induced polarization (second term in Eq.~\ref{eq:p_equation}) is derivable from a functional,  $- (\zeta_{p}/\Gamma_p)\int d^{2}x \, m  \partial_i p_i$, corresponding to an effective free energy for motor-induced spontaneous splay.
 This corresponds to a negative contribution to surface energy \cite{Pettey99}  (see SI IV for the derivation), that drives the droplet shape instability when $\zeta_{p} m_{0}/\Gamma_p >\gamma $  . We thus expect the transition between undeformed and pinched droplets to occur across a dividing line, $\zeta_p \sim \gamma$, which is confirmed by our phase diagram data in Fig.~\ref{fig:combined_fig}(i).

By further increasing $\zeta_p$ and lowering surface tension, we find droplets with two asters which pinch them into  three-lobed structures (Fig.~\ref{fig:combined_fig}(iv), green upward triangles). Importantly, we also find experimental realizations of such multiply pinched, equi-lobed structures, shown in Fig.~\ref{fig:combined_fig}(iv) bottom panel. Such linear ``strings of tactoids'' connected by multiple motor clusters  have not been previously reported and are evocative of fibers with periodic contractile units such as in muscle or anomalous, multipolar biological spindles \cite{Ganem2009}.

Multi-aster states such as aster lattices  are in fact a generic feature of  bulk active polar fluids \cite{Sankararaman2004, gowrishankar2016nonequilibrium}, but here, we report and analyze their occurrence within droplets. To explore the aster-forming instability arising from the feedback between motor flux and motor induced polarization in Eq.~\ref{eq:c_equation} and Eq.~\ref{eq:p_equation}, we perform a linear stability analysis for an incompressible, polar fluid featuring only $p$ and $m$ (see SI V). The analysis yields a characteristic spacing between asters in bulk, $l_c$, such that $l_c \sim 1/\sqrt{m_{0} \zeta_{p}}$, that decreases with motor density and motor induced polarization. Based on this, we expect multiple asters can be accommodated within the droplet when its major axis, $l_{d}$, is long enough compared to $l_c$. 
We confirm this numerically in (Fig.~\ref{fig:combined_fig}(vi)) by showing that the critical motor concentration for transition from one  to two asters for varying  droplet size (but all other parameters constant),
does indeed decrease with initial droplet size, with an expected scaling of $m_{0} \sim 1/l_{d}^{2}$. 
Since the aspect ratio of a droplet at equilibrium increases with decreasing surface tension, the droplets get longer and can accommodate multiple asters when $l_{d} >l_{c} \sim \zeta_{p}^{-1/2}$ is satisfied.  This explains why lower surface tension favors the two-aster state (green triangles) in Fig.~\ref{fig:combined_fig}(i).   
At even higher $\zeta_{p}$ and lower $\gamma$,  we find a region of the phase diagram in Fig.~\ref{fig:combined_fig}(i) where the droplet is fully divided (purple diamonds). 

To see how droplet division can be enhanced, we now explore an alternative shape instability in our model Eq.~\ref{eq:psi_equation}, whereby an active density flux pushes filaments away from the aster. By varying the  strength of this active flux, $\zeta$,  and the surface tension $\gamma$, while keeping $\zeta_p$($=5.0$) fixed, we obtain the phase diagram in Fig.~\ref{fig:combined_fig}(vii), which shows undeformed single tactoids (blue stars), pinched droplets (orange circles) and fully divided droplets (purple diamonds). Here, we find that droplet division occurs over a wide range in parameter space, showing that $\zeta$ is a good control parameter to trigger droplet division, while $\zeta_{p}$ can be used to obtain multi-asters states.

To analyze how the active flux parameter, $\zeta$, for full division scales with $\gamma$, we note that if the motor and polarization are ``fast variables'' that relax quickly to their steady state aster solution (shown in the SI VI), the active density flux term in Eq.~\eqref{eq:psi_equation}, $-\zeta \nabla \cdot(m \bm{p})$ can be 
obtained from the variation of an effective ``free energy'' functional, $-\zeta \int d^{2}x \, \omega \psi$ . Here, $\omega$ is a scalar potential that gives the steady state  polarization corresponding to an aster through $m \bm{p} = -\nabla \omega$.
It can be shown that this effective free energy term makes a negative contribution to the droplet surface energy (detailed in the SI VI) that scales with $\zeta$ and which can then destabilize the droplet when sufficiently strong compared to surface tension. We thus expect the transition to fully divided droplets to scale as $\zeta \sim \gamma$, 
which is confirmed by our simulations in Fig.~\ref{fig:combined_fig}(vii).

\textit{Conclusion}. 
We build and explore a minimal theoretical model for motor--filament droplets which captures four different shapes and show them experimentally in the actomyosin droplet system. We predict a phase diagram of expected droplet shapes based on motor activity and filament interactions, which can be tested in future experiments by systematically varying these properties. Unlike other theoretically proposed active droplet division mechanisms, the droplet shape changes we predict and show in experiment, do not require fluxes of chemical reactants \cite{Zwicker2017} or large-scale fluid flows and defect dynamics \cite{Giomi2014}, that arise in active nematic morphodynamics \cite{Metselaar2019}. Instead, the emergence of local polar order within a nematic and the consequent spatial localization of activity, is crucial in our model, which was not explored in a previous equilibrium nematic model for this system \cite{Weirich19}.
In addition to \emph{in vitro} cytoskeletal spindles, our results also have implications for the physical principles behind cell division and the organization of the biological spindle, where  motor-induced active organization of filaments~\cite{Brugues2014}, shape change \cite{Oriola2020}, co-existence of polar order with nematic order \cite{Roostalu2018} and even formation of multi-lobed structures \cite{Leoni2017} can occur. We expect our work to inform strategies to use localization of active agents to achieve self-actuated shape morphing in materials \cite{Holmes2019} and synthetic cells. 

\begin{acknowledgments}
  FJS and KD gratefully acknowledge computing time on the  Multi-Environment Computer for Exploration and Discovery (MERCED)
  cluster at the University of California, Merced, which was funded by National Science Foundation Grant No. ACI-1429783 and support from the National Science Foundation: NSF-CREST: Center for Cellular and Biomolecular Machines (CCBM) at the University of California, Merced: NSF-HRD-1547848.. PR received funding from the ``Investissements d'Avenir'' French Government program managed by the French National Research Agency (ANR-16-CONV-0001) and from Excellence
Initiative of Aix-Marseille University - A*MIDEX.
\end{acknowledgments}

\bibliography{main}

\begin{thebibliography}{53}%
\makeatletter
\providecommand \@ifxundefined [1]{%
 \@ifx{#1\undefined}
}%
\providecommand \@ifnum [1]{%
 \ifnum #1\expandafter \@firstoftwo
 \else \expandafter \@secondoftwo
 \fi
}%
\providecommand \@ifx [1]{%
 \ifx #1\expandafter \@firstoftwo
 \else \expandafter \@secondoftwo
 \fi
}%
\providecommand \natexlab [1]{#1}%
\providecommand \enquote  [1]{``#1''}%
\providecommand \bibnamefont  [1]{#1}%
\providecommand \bibfnamefont [1]{#1}%
\providecommand \citenamefont [1]{#1}%
\providecommand \href@noop [0]{\@secondoftwo}%
\providecommand \href [0]{\begingroup \@sanitize@url \@href}%
\providecommand \@href[1]{\@@startlink{#1}\@@href}%
\providecommand \@@href[1]{\endgroup#1\@@endlink}%
\providecommand \@sanitize@url [0]{\catcode `\\12\catcode `\$12\catcode
  `\&12\catcode `\#12\catcode `\^12\catcode `\_12\catcode `\%12\relax}%
\providecommand \@@startlink[1]{}%
\providecommand \@@endlink[0]{}%
\providecommand \url  [0]{\begingroup\@sanitize@url \@url }%
\providecommand \@url [1]{\endgroup\@href {#1}{\urlprefix }}%
\providecommand \urlprefix  [0]{URL }%
\providecommand \Eprint [0]{\href }%
\providecommand \doibase [0]{https://doi.org/}%
\providecommand \selectlanguage [0]{\@gobble}%
\providecommand \bibinfo  [0]{\@secondoftwo}%
\providecommand \bibfield  [0]{\@secondoftwo}%
\providecommand \translation [1]{[#1]}%
\providecommand \BibitemOpen [0]{}%
\providecommand \bibitemStop [0]{}%
\providecommand \bibitemNoStop [0]{.\EOS\space}%
\providecommand \EOS [0]{\spacefactor3000\relax}%
\providecommand \BibitemShut  [1]{\csname bibitem#1\endcsname}%
\let\auto@bib@innerbib\@empty
%</preamble>
\bibitem [{\citenamefont {Murrell}\ \emph {et~al.}(2015)\citenamefont
  {Murrell}, \citenamefont {Oakes}, \citenamefont {Lenz},\ and\ \citenamefont
  {Gardel}}]{gardel_15}%
  \BibitemOpen
  \bibfield  {author} {\bibinfo {author} {\bibfnamefont {M.}~\bibnamefont
  {Murrell}}, \bibinfo {author} {\bibfnamefont {P.~W.}\ \bibnamefont {Oakes}},
  \bibinfo {author} {\bibfnamefont {M.}~\bibnamefont {Lenz}},\ and\ \bibinfo
  {author} {\bibfnamefont {M.~L.}\ \bibnamefont {Gardel}},\ }\bibfield  {title}
  {\bibinfo {title} {Forcing cells into shape: the mechanics of actomyosin
  contractility},\ }\href {https://doi.org/10.1038/nrm4012} {\bibfield
  {journal} {\bibinfo  {journal} {Nature Reviews Molecular Cell Biology}\
  }\textbf {\bibinfo {volume} {16}},\ \bibinfo {pages} {486 EP } (\bibinfo
  {year} {2015})},\ \bibinfo {note} {review Article}\BibitemShut {NoStop}%
\bibitem [{\citenamefont {Noireaux}\ \emph {et~al.}(2011)\citenamefont
  {Noireaux}, \citenamefont {Maeda},\ and\ \citenamefont
  {Libchaber}}]{Noireaux2011}%
  \BibitemOpen
  \bibfield  {author} {\bibinfo {author} {\bibfnamefont {V.}~\bibnamefont
  {Noireaux}}, \bibinfo {author} {\bibfnamefont {Y.~T.}\ \bibnamefont
  {Maeda}},\ and\ \bibinfo {author} {\bibfnamefont {A.}~\bibnamefont
  {Libchaber}},\ }\bibfield  {title} {\bibinfo {title} {Development of an
  artificial cell, from self-organization to computation and
  self-reproduction},\ }\href {https://doi.org/10.1073/pnas.1017075108}
  {\bibfield  {journal} {\bibinfo  {journal} {Proceedings of the National
  Academy of Sciences}\ }\textbf {\bibinfo {volume} {108}},\ \bibinfo {pages}
  {3473} (\bibinfo {year} {2011})},\ \Eprint
  {https://arxiv.org/abs/https://www.pnas.org/content/108/9/3473.full.pdf}
  {https://www.pnas.org/content/108/9/3473.full.pdf} \BibitemShut {NoStop}%
\bibitem [{\citenamefont {Soares~e Silva}\ \emph {et~al.}(2011)\citenamefont
  {Soares~e Silva}, \citenamefont {Depken}, \citenamefont {Stuhrmann},
  \citenamefont {Korsten}, \citenamefont {MacKintosh},\ and\ \citenamefont
  {Koenderink}}]{SoareseSilva2011}%
  \BibitemOpen
  \bibfield  {author} {\bibinfo {author} {\bibfnamefont {M.}~\bibnamefont
  {Soares~e Silva}}, \bibinfo {author} {\bibfnamefont {M.}~\bibnamefont
  {Depken}}, \bibinfo {author} {\bibfnamefont {B.}~\bibnamefont {Stuhrmann}},
  \bibinfo {author} {\bibfnamefont {M.}~\bibnamefont {Korsten}}, \bibinfo
  {author} {\bibfnamefont {F.~C.}\ \bibnamefont {MacKintosh}},\ and\ \bibinfo
  {author} {\bibfnamefont {G.~H.}\ \bibnamefont {Koenderink}},\ }\bibfield
  {title} {\bibinfo {title} {Active multistage coarsening of actin networks
  driven by myosin motors},\ }\href {https://doi.org/10.1073/pnas.1016616108}
  {\bibfield  {journal} {\bibinfo  {journal} {Proceedings of the National
  Academy of Sciences of the United States of America}\ }\textbf {\bibinfo
  {volume} {108}},\ \bibinfo {pages} {9408} (\bibinfo {year} {2011})},\
  \bibinfo {note} {21593409[pmid]}\BibitemShut {NoStop}%
\bibitem [{\citenamefont {Sanchez}\ \emph {et~al.}(2012)\citenamefont
  {Sanchez}, \citenamefont {Chen}, \citenamefont {DeCamp}, \citenamefont
  {Heymann},\ and\ \citenamefont {Dogic}}]{sanchez2012spontaneous}%
  \BibitemOpen
  \bibfield  {author} {\bibinfo {author} {\bibfnamefont {T.}~\bibnamefont
  {Sanchez}}, \bibinfo {author} {\bibfnamefont {D.~T.}\ \bibnamefont {Chen}},
  \bibinfo {author} {\bibfnamefont {S.~J.}\ \bibnamefont {DeCamp}}, \bibinfo
  {author} {\bibfnamefont {M.}~\bibnamefont {Heymann}},\ and\ \bibinfo {author}
  {\bibfnamefont {Z.}~\bibnamefont {Dogic}},\ }\bibfield  {title} {\bibinfo
  {title} {Spontaneous motion in hierarchically assembled active matter},\
  }\href@noop {} {\bibfield  {journal} {\bibinfo  {journal} {Nature}\ }\textbf
  {\bibinfo {volume} {491}},\ \bibinfo {pages} {431} (\bibinfo {year}
  {2012})}\BibitemShut {NoStop}%
\bibitem [{\citenamefont {Marchetti}\ \emph {et~al.}(2013)\citenamefont
  {Marchetti}, \citenamefont {Joanny}, \citenamefont {Ramaswamy}, \citenamefont
  {Liverpool}, \citenamefont {Prost}, \citenamefont {Rao},\ and\ \citenamefont
  {Simha}}]{Marchetti2013}%
  \BibitemOpen
  \bibfield  {author} {\bibinfo {author} {\bibfnamefont {M.~C.}\ \bibnamefont
  {Marchetti}}, \bibinfo {author} {\bibfnamefont {J.~F.}\ \bibnamefont
  {Joanny}}, \bibinfo {author} {\bibfnamefont {S.}~\bibnamefont {Ramaswamy}},
  \bibinfo {author} {\bibfnamefont {T.~B.}\ \bibnamefont {Liverpool}}, \bibinfo
  {author} {\bibfnamefont {J.}~\bibnamefont {Prost}}, \bibinfo {author}
  {\bibfnamefont {M.}~\bibnamefont {Rao}},\ and\ \bibinfo {author}
  {\bibfnamefont {R.~A.}\ \bibnamefont {Simha}},\ }\bibfield  {title} {\bibinfo
  {title} {Hydrodynamics of soft active matter},\ }\href
  {https://doi.org/10.1103/RevModPhys.85.1143} {\bibfield  {journal} {\bibinfo
  {journal} {Rev. Mod. Phys.}\ }\textbf {\bibinfo {volume} {85}},\ \bibinfo
  {pages} {1143} (\bibinfo {year} {2013})}\BibitemShut {NoStop}%
\bibitem [{\citenamefont {Hyman}\ \emph {et~al.}(2014)\citenamefont {Hyman},
  \citenamefont {Weber},\ and\ \citenamefont {Jülicher}}]{Hyman2014}%
  \BibitemOpen
  \bibfield  {author} {\bibinfo {author} {\bibfnamefont {A.~A.}\ \bibnamefont
  {Hyman}}, \bibinfo {author} {\bibfnamefont {C.~A.}\ \bibnamefont {Weber}},\
  and\ \bibinfo {author} {\bibfnamefont {F.}~\bibnamefont {Jülicher}},\
  }\bibfield  {title} {\bibinfo {title} {Liquid-liquid phase separation in
  biology},\ }\href {https://doi.org/10.1146/annurev-cellbio-100913-013325}
  {\bibfield  {journal} {\bibinfo  {journal} {Annual Review of Cell and
  Developmental Biology}\ }\textbf {\bibinfo {volume} {30}},\ \bibinfo {pages}
  {39} (\bibinfo {year} {2014})},\ \bibinfo {note} {pMID: 25288112},\ \Eprint
  {https://arxiv.org/abs/https://doi.org/10.1146/annurev-cellbio-100913-013325}
  {https://doi.org/10.1146/annurev-cellbio-100913-013325} \BibitemShut
  {NoStop}%
\bibitem [{\citenamefont {De~Gennes}\ and\ \citenamefont
  {Prost}(1995)}]{DeGennesText}%
  \BibitemOpen
  \bibfield  {author} {\bibinfo {author} {\bibfnamefont {P.-G.}\ \bibnamefont
  {De~Gennes}}\ and\ \bibinfo {author} {\bibfnamefont {J.}~\bibnamefont
  {Prost}},\ }\href@noop {} {\emph {\bibinfo {title} {The Physics of Liquid
  Crystals}}},\ Vol.~\bibinfo {volume} {83}\ (\bibinfo  {publisher} {Oxford
  University Press},\ \bibinfo {year} {1995})\BibitemShut {NoStop}%
\bibitem [{\citenamefont {Weirich}\ \emph {et~al.}(2017)\citenamefont
  {Weirich}, \citenamefont {Banerjee}, \citenamefont {Dasbiswas}, \citenamefont
  {Witten}, \citenamefont {Vaikuntanathan},\ and\ \citenamefont
  {Gardel}}]{Weirich17}%
  \BibitemOpen
  \bibfield  {author} {\bibinfo {author} {\bibfnamefont {K.~L.}\ \bibnamefont
  {Weirich}}, \bibinfo {author} {\bibfnamefont {S.}~\bibnamefont {Banerjee}},
  \bibinfo {author} {\bibfnamefont {K.}~\bibnamefont {Dasbiswas}}, \bibinfo
  {author} {\bibfnamefont {T.~A.}\ \bibnamefont {Witten}}, \bibinfo {author}
  {\bibfnamefont {S.}~\bibnamefont {Vaikuntanathan}},\ and\ \bibinfo {author}
  {\bibfnamefont {M.~L.}\ \bibnamefont {Gardel}},\ }\bibfield  {title}
  {\bibinfo {title} {Liquid behavior of cross-linked actin bundles},\ }\href
  {https://doi.org/10.1073/pnas.1616133114} {\bibfield  {journal} {\bibinfo
  {journal} {Proceedings of the National Academy of Sciences}\ }\textbf
  {\bibinfo {volume} {114}},\ \bibinfo {pages} {2131} (\bibinfo {year}
  {2017})}\BibitemShut {NoStop}%
\bibitem [{\citenamefont {Brugu{\'e}s}\ and\ \citenamefont
  {Needleman}(2014)}]{Brugues2014}%
  \BibitemOpen
  \bibfield  {author} {\bibinfo {author} {\bibfnamefont {J.}~\bibnamefont
  {Brugu{\'e}s}}\ and\ \bibinfo {author} {\bibfnamefont {D.}~\bibnamefont
  {Needleman}},\ }\bibfield  {title} {\bibinfo {title} {Physical basis of
  spindle self-organization},\ }\href {https://doi.org/10.1073/pnas.1409404111}
  {\bibfield  {journal} {\bibinfo  {journal} {Proceedings of the National
  Academy of Sciences}\ }\textbf {\bibinfo {volume} {111}},\ \bibinfo {pages}
  {18496} (\bibinfo {year} {2014})},\ \Eprint
  {https://arxiv.org/abs/https://www.pnas.org/content/111/52/18496.full.pdf}
  {https://www.pnas.org/content/111/52/18496.full.pdf} \BibitemShut {NoStop}%
\bibitem [{\citenamefont {Kaznacheev}\ \emph {et~al.}(2002)\citenamefont
  {Kaznacheev}, \citenamefont {Bogdanov},\ and\ \citenamefont
  {Taraskin}}]{Taraskin02}%
  \BibitemOpen
  \bibfield  {author} {\bibinfo {author} {\bibfnamefont {A.~V.}\ \bibnamefont
  {Kaznacheev}}, \bibinfo {author} {\bibfnamefont {M.~M.}\ \bibnamefont
  {Bogdanov}},\ and\ \bibinfo {author} {\bibfnamefont {S.~A.}\ \bibnamefont
  {Taraskin}},\ }\bibfield  {title} {\bibinfo {title} {The nature of prolate
  shape of tactoids in lyotropic inorganic liquid crystals},\ }\href
  {https://doi.org/10.1134/1.1499901} {\bibfield  {journal} {\bibinfo
  {journal} {Journal of Experimental and Theoretical Physics}\ }\textbf
  {\bibinfo {volume} {95}},\ \bibinfo {pages} {57} (\bibinfo {year}
  {2002})}\BibitemShut {NoStop}%
\bibitem [{\citenamefont {Prinsen}\ and\ \citenamefont {van~der
  Schoot}(2003)}]{vdS03}%
  \BibitemOpen
  \bibfield  {author} {\bibinfo {author} {\bibfnamefont {P.}~\bibnamefont
  {Prinsen}}\ and\ \bibinfo {author} {\bibfnamefont {P.}~\bibnamefont {van~der
  Schoot}},\ }\bibfield  {title} {\bibinfo {title} {Shape and director-field
  transformation of tactoids},\ }\href
  {https://doi.org/10.1103/PhysRevE.68.021701} {\bibfield  {journal} {\bibinfo
  {journal} {Phys. Rev. E}\ }\textbf {\bibinfo {volume} {68}},\ \bibinfo
  {pages} {021701} (\bibinfo {year} {2003})}\BibitemShut {NoStop}%
\bibitem [{\citenamefont {Giomi}\ and\ \citenamefont
  {DeSimone}(2014)}]{Giomi2014}%
  \BibitemOpen
  \bibfield  {author} {\bibinfo {author} {\bibfnamefont {L.}~\bibnamefont
  {Giomi}}\ and\ \bibinfo {author} {\bibfnamefont {A.}~\bibnamefont
  {DeSimone}},\ }\bibfield  {title} {\bibinfo {title} {Spontaneous division and
  motility in active nematic droplets},\ }\href
  {https://doi.org/10.1103/PhysRevLett.112.147802} {\bibfield  {journal}
  {\bibinfo  {journal} {Phys. Rev. Lett.}\ }\textbf {\bibinfo {volume} {112}},\
  \bibinfo {pages} {147802} (\bibinfo {year} {2014})}\BibitemShut {NoStop}%
\bibitem [{\citenamefont {Zwicker}\ \emph {et~al.}(2017)\citenamefont
  {Zwicker}, \citenamefont {Seyboldt}, \citenamefont {Weber}, \citenamefont
  {Hyman},\ and\ \citenamefont {J{\"u}licher}}]{Zwicker2017}%
  \BibitemOpen
  \bibfield  {author} {\bibinfo {author} {\bibfnamefont {D.}~\bibnamefont
  {Zwicker}}, \bibinfo {author} {\bibfnamefont {R.}~\bibnamefont {Seyboldt}},
  \bibinfo {author} {\bibfnamefont {C.~A.}\ \bibnamefont {Weber}}, \bibinfo
  {author} {\bibfnamefont {A.~A.}\ \bibnamefont {Hyman}},\ and\ \bibinfo
  {author} {\bibfnamefont {F.}~\bibnamefont {J{\"u}licher}},\ }\bibfield
  {title} {\bibinfo {title} {Growth and division of active droplets provides a
  model for protocells},\ }\href {https://doi.org/10.1038/nphys3984} {\bibfield
   {journal} {\bibinfo  {journal} {Nature Physics}\ }\textbf {\bibinfo {volume}
  {13}},\ \bibinfo {pages} {408} (\bibinfo {year} {2017})}\BibitemShut
  {NoStop}%
\bibitem [{\citenamefont {Vutukuri}\ \emph {et~al.}(2020)\citenamefont
  {Vutukuri}, \citenamefont {Hoore}, \citenamefont {Abaurrea-Velasco},
  \citenamefont {van Buren}, \citenamefont {Dutto}, \citenamefont {Auth},
  \citenamefont {Fedosov}, \citenamefont {Gompper},\ and\ \citenamefont
  {Vermant}}]{vutukuri2020}%
  \BibitemOpen
  \bibfield  {author} {\bibinfo {author} {\bibfnamefont {H.~R.}\ \bibnamefont
  {Vutukuri}}, \bibinfo {author} {\bibfnamefont {M.}~\bibnamefont {Hoore}},
  \bibinfo {author} {\bibfnamefont {C.}~\bibnamefont {Abaurrea-Velasco}},
  \bibinfo {author} {\bibfnamefont {L.}~\bibnamefont {van Buren}}, \bibinfo
  {author} {\bibfnamefont {A.}~\bibnamefont {Dutto}}, \bibinfo {author}
  {\bibfnamefont {T.}~\bibnamefont {Auth}}, \bibinfo {author} {\bibfnamefont
  {D.~A.}\ \bibnamefont {Fedosov}}, \bibinfo {author} {\bibfnamefont
  {G.}~\bibnamefont {Gompper}},\ and\ \bibinfo {author} {\bibfnamefont
  {J.}~\bibnamefont {Vermant}},\ }\bibfield  {title} {\bibinfo {title} {Active
  particles induce large shape deformations in giant lipid vesicles},\
  }\href@noop {} {\bibfield  {journal} {\bibinfo  {journal} {Nature}\ }\textbf
  {\bibinfo {volume} {586}},\ \bibinfo {pages} {52} (\bibinfo {year}
  {2020})}\BibitemShut {NoStop}%
\bibitem [{\citenamefont {Wang}\ \emph {et~al.}(2019)\citenamefont {Wang},
  \citenamefont {Guo}, \citenamefont {Tian},\ and\ \citenamefont
  {Chen}}]{wang2019}%
  \BibitemOpen
  \bibfield  {author} {\bibinfo {author} {\bibfnamefont {C.}~\bibnamefont
  {Wang}}, \bibinfo {author} {\bibfnamefont {Y.-k.}\ \bibnamefont {Guo}},
  \bibinfo {author} {\bibfnamefont {W.-d.}\ \bibnamefont {Tian}},\ and\
  \bibinfo {author} {\bibfnamefont {K.}~\bibnamefont {Chen}},\ }\bibfield
  {title} {\bibinfo {title} {Shape transformation and manipulation of a vesicle
  by active particles},\ }\href@noop {} {\bibfield  {journal} {\bibinfo
  {journal} {The Journal of chemical physics}\ }\textbf {\bibinfo {volume}
  {150}},\ \bibinfo {pages} {044907} (\bibinfo {year} {2019})}\BibitemShut
  {NoStop}%
\bibitem [{\citenamefont {Li}\ and\ \citenamefont {ten Wolde}(2019)}]{li2019}%
  \BibitemOpen
  \bibfield  {author} {\bibinfo {author} {\bibfnamefont {Y.}~\bibnamefont
  {Li}}\ and\ \bibinfo {author} {\bibfnamefont {P.~R.}\ \bibnamefont {ten
  Wolde}},\ }\bibfield  {title} {\bibinfo {title} {Shape transformations of
  vesicles induced by swim pressure},\ }\href@noop {} {\bibfield  {journal}
  {\bibinfo  {journal} {Physical review letters}\ }\textbf {\bibinfo {volume}
  {123}},\ \bibinfo {pages} {148003} (\bibinfo {year} {2019})}\BibitemShut
  {NoStop}%
\bibitem [{\citenamefont {Takatori}\ and\ \citenamefont
  {Sahu}(2020)}]{takatori2020}%
  \BibitemOpen
  \bibfield  {author} {\bibinfo {author} {\bibfnamefont {S.~C.}\ \bibnamefont
  {Takatori}}\ and\ \bibinfo {author} {\bibfnamefont {A.}~\bibnamefont
  {Sahu}},\ }\bibfield  {title} {\bibinfo {title} {Active contact forces drive
  nonequilibrium fluctuations in membrane vesicles},\ }\href@noop {} {\bibfield
   {journal} {\bibinfo  {journal} {Physical Review Letters}\ }\textbf {\bibinfo
  {volume} {124}},\ \bibinfo {pages} {158102} (\bibinfo {year}
  {2020})}\BibitemShut {NoStop}%
\bibitem [{\citenamefont {Singh}\ and\ \citenamefont
  {Cates}(2019)}]{singh2019}%
  \BibitemOpen
  \bibfield  {author} {\bibinfo {author} {\bibfnamefont {R.}~\bibnamefont
  {Singh}}\ and\ \bibinfo {author} {\bibfnamefont {M.}~\bibnamefont {Cates}},\
  }\bibfield  {title} {\bibinfo {title} {Hydrodynamically interrupted droplet
  growth in scalar active matter},\ }\href@noop {} {\bibfield  {journal}
  {\bibinfo  {journal} {Physical review letters}\ }\textbf {\bibinfo {volume}
  {123}},\ \bibinfo {pages} {148005} (\bibinfo {year} {2019})}\BibitemShut
  {NoStop}%
\bibitem [{\citenamefont {Ruiz-Herrero}\ \emph {et~al.}(2019)\citenamefont
  {Ruiz-Herrero}, \citenamefont {Fai},\ and\ \citenamefont
  {Mahadevan}}]{ruiz2019}%
  \BibitemOpen
  \bibfield  {author} {\bibinfo {author} {\bibfnamefont {T.}~\bibnamefont
  {Ruiz-Herrero}}, \bibinfo {author} {\bibfnamefont {T.~G.}\ \bibnamefont
  {Fai}},\ and\ \bibinfo {author} {\bibfnamefont {L.}~\bibnamefont
  {Mahadevan}},\ }\bibfield  {title} {\bibinfo {title} {Dynamics of growth and
  form in prebiotic vesicles},\ }\href@noop {} {\bibfield  {journal} {\bibinfo
  {journal} {Physical review letters}\ }\textbf {\bibinfo {volume} {123}},\
  \bibinfo {pages} {038102} (\bibinfo {year} {2019})}\BibitemShut {NoStop}%
\bibitem [{\citenamefont {Tjhung}\ \emph {et~al.}(2012)\citenamefont {Tjhung},
  \citenamefont {Marenduzzo},\ and\ \citenamefont {Cates}}]{Tjhung2012}%
  \BibitemOpen
  \bibfield  {author} {\bibinfo {author} {\bibfnamefont {E.}~\bibnamefont
  {Tjhung}}, \bibinfo {author} {\bibfnamefont {D.}~\bibnamefont {Marenduzzo}},\
  and\ \bibinfo {author} {\bibfnamefont {M.~E.}\ \bibnamefont {Cates}},\
  }\bibfield  {title} {\bibinfo {title} {Spontaneous symmetry breaking in
  active droplets provides a generic route to motility},\ }\href
  {https://doi.org/10.1073/pnas.1200843109} {\bibfield  {journal} {\bibinfo
  {journal} {Proceedings of the National Academy of Sciences}\ }\textbf
  {\bibinfo {volume} {109}},\ \bibinfo {pages} {12381} (\bibinfo {year}
  {2012})},\ \Eprint
  {https://arxiv.org/abs/https://www.pnas.org/content/109/31/12381.full.pdf}
  {https://www.pnas.org/content/109/31/12381.full.pdf} \BibitemShut {NoStop}%
\bibitem [{\citenamefont {Ziebert}\ \emph {et~al.}(2012)\citenamefont
  {Ziebert}, \citenamefont {Swaminathan},\ and\ \citenamefont
  {Aranson}}]{ziebert2012}%
  \BibitemOpen
  \bibfield  {author} {\bibinfo {author} {\bibfnamefont {F.}~\bibnamefont
  {Ziebert}}, \bibinfo {author} {\bibfnamefont {S.}~\bibnamefont
  {Swaminathan}},\ and\ \bibinfo {author} {\bibfnamefont {I.~S.}\ \bibnamefont
  {Aranson}},\ }\bibfield  {title} {\bibinfo {title} {Model for
  self-polarization and motility of keratocyte fragments},\ }\href@noop {}
  {\bibfield  {journal} {\bibinfo  {journal} {Journal of The Royal Society
  Interface}\ }\textbf {\bibinfo {volume} {9}},\ \bibinfo {pages} {1084}
  (\bibinfo {year} {2012})}\BibitemShut {NoStop}%
\bibitem [{\citenamefont {Keber}\ \emph {et~al.}(2014)\citenamefont {Keber},
  \citenamefont {Loiseau}, \citenamefont {Sanchez}, \citenamefont {DeCamp},
  \citenamefont {Giomi}, \citenamefont {Bowick}, \citenamefont {Marchetti},
  \citenamefont {Dogic},\ and\ \citenamefont {Bausch}}]{Keber2014}%
  \BibitemOpen
  \bibfield  {author} {\bibinfo {author} {\bibfnamefont {F.~C.}\ \bibnamefont
  {Keber}}, \bibinfo {author} {\bibfnamefont {E.}~\bibnamefont {Loiseau}},
  \bibinfo {author} {\bibfnamefont {T.}~\bibnamefont {Sanchez}}, \bibinfo
  {author} {\bibfnamefont {S.~J.}\ \bibnamefont {DeCamp}}, \bibinfo {author}
  {\bibfnamefont {L.}~\bibnamefont {Giomi}}, \bibinfo {author} {\bibfnamefont
  {M.~J.}\ \bibnamefont {Bowick}}, \bibinfo {author} {\bibfnamefont {M.~C.}\
  \bibnamefont {Marchetti}}, \bibinfo {author} {\bibfnamefont {Z.}~\bibnamefont
  {Dogic}},\ and\ \bibinfo {author} {\bibfnamefont {A.~R.}\ \bibnamefont
  {Bausch}},\ }\bibfield  {title} {\bibinfo {title} {Topology and dynamics of
  active nematic vesicles},\ }\href@noop {} {\bibfield  {journal} {\bibinfo
  {journal} {Science}\ }\textbf {\bibinfo {volume} {345}},\ \bibinfo {pages}
  {1135} (\bibinfo {year} {2014})}\BibitemShut {NoStop}%
\bibitem [{\citenamefont {Surrey}\ \emph {et~al.}(2001)\citenamefont {Surrey},
  \citenamefont {N{\'e}d{\'e}lec}, \citenamefont {Leibler},\ and\ \citenamefont
  {Karsenti}}]{Surrey2001}%
  \BibitemOpen
  \bibfield  {author} {\bibinfo {author} {\bibfnamefont {T.}~\bibnamefont
  {Surrey}}, \bibinfo {author} {\bibfnamefont {F.}~\bibnamefont
  {N{\'e}d{\'e}lec}}, \bibinfo {author} {\bibfnamefont {S.}~\bibnamefont
  {Leibler}},\ and\ \bibinfo {author} {\bibfnamefont {E.}~\bibnamefont
  {Karsenti}},\ }\bibfield  {title} {\bibinfo {title} {Physical properties
  determining self-organization of motors and microtubules},\ }\href@noop {}
  {\bibfield  {journal} {\bibinfo  {journal} {Science}\ }\textbf {\bibinfo
  {volume} {292}},\ \bibinfo {pages} {1167} (\bibinfo {year}
  {2001})}\BibitemShut {NoStop}%
\bibitem [{\citenamefont {Youn~Lee}\ and\ \citenamefont
  {Kardar}(2001)}]{Lee2001}%
  \BibitemOpen
  \bibfield  {author} {\bibinfo {author} {\bibfnamefont {H.}~\bibnamefont
  {Youn~Lee}}\ and\ \bibinfo {author} {\bibfnamefont {M.}~\bibnamefont
  {Kardar}},\ }\bibfield  {title} {\bibinfo {title} {Macroscopic equations for
  pattern formation in mixtures of microtubules and molecular motors},\ }\href
  {https://doi.org/10.1103/PhysRevE.64.056113} {\bibfield  {journal} {\bibinfo
  {journal} {Phys. Rev. E}\ }\textbf {\bibinfo {volume} {64}},\ \bibinfo
  {pages} {056113} (\bibinfo {year} {2001})}\BibitemShut {NoStop}%
\bibitem [{\citenamefont {Kruse}\ \emph {et~al.}(2004)\citenamefont {Kruse},
  \citenamefont {Joanny}, \citenamefont {J\"ulicher}, \citenamefont {Prost},\
  and\ \citenamefont {Sekimoto}}]{Kruse2004}%
  \BibitemOpen
  \bibfield  {author} {\bibinfo {author} {\bibfnamefont {K.}~\bibnamefont
  {Kruse}}, \bibinfo {author} {\bibfnamefont {J.~F.}\ \bibnamefont {Joanny}},
  \bibinfo {author} {\bibfnamefont {F.}~\bibnamefont {J\"ulicher}}, \bibinfo
  {author} {\bibfnamefont {J.}~\bibnamefont {Prost}},\ and\ \bibinfo {author}
  {\bibfnamefont {K.}~\bibnamefont {Sekimoto}},\ }\bibfield  {title} {\bibinfo
  {title} {Asters, vortices, and rotating spirals in active gels of polar
  filaments},\ }\href {https://doi.org/10.1103/PhysRevLett.92.078101}
  {\bibfield  {journal} {\bibinfo  {journal} {Phys. Rev. Lett.}\ }\textbf
  {\bibinfo {volume} {92}},\ \bibinfo {pages} {078101} (\bibinfo {year}
  {2004})}\BibitemShut {NoStop}%
\bibitem [{\citenamefont {Sankararaman}\ \emph {et~al.}(2004)\citenamefont
  {Sankararaman}, \citenamefont {Menon},\ and\ \citenamefont
  {Sunil~Kumar}}]{Sankararaman2004}%
  \BibitemOpen
  \bibfield  {author} {\bibinfo {author} {\bibfnamefont {S.}~\bibnamefont
  {Sankararaman}}, \bibinfo {author} {\bibfnamefont {G.~I.}\ \bibnamefont
  {Menon}},\ and\ \bibinfo {author} {\bibfnamefont {P.~B.}\ \bibnamefont
  {Sunil~Kumar}},\ }\bibfield  {title} {\bibinfo {title} {Self-organized
  pattern formation in motor-microtubule mixtures},\ }\href
  {https://doi.org/10.1103/PhysRevE.70.031905} {\bibfield  {journal} {\bibinfo
  {journal} {Phys. Rev. E}\ }\textbf {\bibinfo {volume} {70}},\ \bibinfo
  {pages} {031905} (\bibinfo {year} {2004})}\BibitemShut {NoStop}%
\bibitem [{\citenamefont {Aranson}\ and\ \citenamefont
  {Tsimring}(2006)}]{Aranson2006}%
  \BibitemOpen
  \bibfield  {author} {\bibinfo {author} {\bibfnamefont {I.~S.}\ \bibnamefont
  {Aranson}}\ and\ \bibinfo {author} {\bibfnamefont {L.~S.}\ \bibnamefont
  {Tsimring}},\ }\bibfield  {title} {\bibinfo {title} {Theory of self-assembly
  of microtubules and motors},\ }\href
  {https://doi.org/10.1103/PhysRevE.74.031915} {\bibfield  {journal} {\bibinfo
  {journal} {Phys. Rev. E}\ }\textbf {\bibinfo {volume} {74}},\ \bibinfo
  {pages} {031915} (\bibinfo {year} {2006})}\BibitemShut {NoStop}%
\bibitem [{\citenamefont {Ahmadi}\ \emph {et~al.}(2005)\citenamefont {Ahmadi},
  \citenamefont {Liverpool},\ and\ \citenamefont {Marchetti}}]{Ahmadi2005}%
  \BibitemOpen
  \bibfield  {author} {\bibinfo {author} {\bibfnamefont {A.}~\bibnamefont
  {Ahmadi}}, \bibinfo {author} {\bibfnamefont {T.~B.}\ \bibnamefont
  {Liverpool}},\ and\ \bibinfo {author} {\bibfnamefont {M.~C.}\ \bibnamefont
  {Marchetti}},\ }\bibfield  {title} {\bibinfo {title} {Nematic and polar order
  in active filament solutions},\ }\href
  {https://doi.org/10.1103/PhysRevE.72.060901} {\bibfield  {journal} {\bibinfo
  {journal} {Phys. Rev. E}\ }\textbf {\bibinfo {volume} {72}},\ \bibinfo
  {pages} {060901} (\bibinfo {year} {2005})}\BibitemShut {NoStop}%
\bibitem [{\citenamefont {Ziebert}\ and\ \citenamefont
  {Zimmermann}(2005)}]{Ziebert2005}%
  \BibitemOpen
  \bibfield  {author} {\bibinfo {author} {\bibfnamefont {F.}~\bibnamefont
  {Ziebert}}\ and\ \bibinfo {author} {\bibfnamefont {W.}~\bibnamefont
  {Zimmermann}},\ }\bibfield  {title} {\bibinfo {title} {Nonlinear competition
  between asters and stripes in filament-motor systems},\ }\href
  {https://doi.org/10.1140/epje/i2005-10029-3} {\bibfield  {journal} {\bibinfo
  {journal} {The European Physical Journal E}\ }\textbf {\bibinfo {volume}
  {18}},\ \bibinfo {pages} {41} (\bibinfo {year} {2005})}\BibitemShut {NoStop}%
\bibitem [{\citenamefont {Gowrishankar}\ and\ \citenamefont
  {Rao}(2016)}]{gowrishankar2016nonequilibrium}%
  \BibitemOpen
  \bibfield  {author} {\bibinfo {author} {\bibfnamefont {K.}~\bibnamefont
  {Gowrishankar}}\ and\ \bibinfo {author} {\bibfnamefont {M.}~\bibnamefont
  {Rao}},\ }\bibfield  {title} {\bibinfo {title} {Nonequilibrium phase
  transitions, fluctuations and correlations in an active contractile polar
  fluid},\ }\href@noop {} {\bibfield  {journal} {\bibinfo  {journal} {Soft
  matter}\ }\textbf {\bibinfo {volume} {12}},\ \bibinfo {pages} {2040}
  (\bibinfo {year} {2016})}\BibitemShut {NoStop}%
\bibitem [{\citenamefont {Weirich}\ \emph {et~al.}(2019)\citenamefont
  {Weirich}, \citenamefont {Dasbiswas}, \citenamefont {Witten}, \citenamefont
  {Vaikuntanathan},\ and\ \citenamefont {Gardel}}]{Weirich19}%
  \BibitemOpen
  \bibfield  {author} {\bibinfo {author} {\bibfnamefont {K.~L.}\ \bibnamefont
  {Weirich}}, \bibinfo {author} {\bibfnamefont {K.}~\bibnamefont {Dasbiswas}},
  \bibinfo {author} {\bibfnamefont {T.~A.}\ \bibnamefont {Witten}}, \bibinfo
  {author} {\bibfnamefont {S.}~\bibnamefont {Vaikuntanathan}},\ and\ \bibinfo
  {author} {\bibfnamefont {M.~L.}\ \bibnamefont {Gardel}},\ }\bibfield  {title}
  {\bibinfo {title} {Self-organizing motors divide active liquid droplets},\
  }\href {https://doi.org/10.1073/pnas.1814854116} {\bibfield  {journal}
  {\bibinfo  {journal} {Proceedings of the National Academy of Sciences}\
  }\textbf {\bibinfo {volume} {116}},\ \bibinfo {pages} {11125} (\bibinfo
  {year} {2019})}\BibitemShut {NoStop}%
\bibitem [{\citenamefont {Kruse}\ and\ \citenamefont
  {J\"ulicher}(2003)}]{Kruse2003}%
  \BibitemOpen
  \bibfield  {author} {\bibinfo {author} {\bibfnamefont {K.}~\bibnamefont
  {Kruse}}\ and\ \bibinfo {author} {\bibfnamefont {F.}~\bibnamefont
  {J\"ulicher}},\ }\bibfield  {title} {\bibinfo {title} {Self-organization and
  mechanical properties of active filament bundles},\ }\href
  {https://doi.org/10.1103/PhysRevE.67.051913} {\bibfield  {journal} {\bibinfo
  {journal} {Phys. Rev. E}\ }\textbf {\bibinfo {volume} {67}},\ \bibinfo
  {pages} {051913} (\bibinfo {year} {2003})}\BibitemShut {NoStop}%
\bibitem [{\citenamefont {Stachowiak}\ \emph {et~al.}(2012)\citenamefont
  {Stachowiak}, \citenamefont {McCall}, \citenamefont {Thoresen}, \citenamefont
  {Balcioglu}, \citenamefont {Kasiewicz}, \citenamefont {Gardel},\ and\
  \citenamefont {O'Shaughnessy}}]{Stachowiak2012}%
  \BibitemOpen
  \bibfield  {author} {\bibinfo {author} {\bibfnamefont {M.}~\bibnamefont
  {Stachowiak}}, \bibinfo {author} {\bibfnamefont {P.}~\bibnamefont {McCall}},
  \bibinfo {author} {\bibfnamefont {T.}~\bibnamefont {Thoresen}}, \bibinfo
  {author} {\bibfnamefont {H.}~\bibnamefont {Balcioglu}}, \bibinfo {author}
  {\bibfnamefont {L.}~\bibnamefont {Kasiewicz}}, \bibinfo {author}
  {\bibfnamefont {M.}~\bibnamefont {Gardel}},\ and\ \bibinfo {author}
  {\bibfnamefont {B.}~\bibnamefont {O'Shaughnessy}},\ }\bibfield  {title}
  {\bibinfo {title} {Self-organization of myosin ii in reconstituted actomyosin
  bundles},\ }\href {https://doi.org/https://doi.org/10.1016/j.bpj.2012.08.028}
  {\bibfield  {journal} {\bibinfo  {journal} {Biophysical Journal}\ }\textbf
  {\bibinfo {volume} {103}},\ \bibinfo {pages} {1265 } (\bibinfo {year}
  {2012})}\BibitemShut {NoStop}%
\bibitem [{\citenamefont {Kumar}\ \emph {et~al.}(2018)\citenamefont {Kumar},
  \citenamefont {Zhang}, \citenamefont {de~Pablo},\ and\ \citenamefont
  {Gardel}}]{Kumar2018}%
  \BibitemOpen
  \bibfield  {author} {\bibinfo {author} {\bibfnamefont {N.}~\bibnamefont
  {Kumar}}, \bibinfo {author} {\bibfnamefont {R.}~\bibnamefont {Zhang}},
  \bibinfo {author} {\bibfnamefont {J.~J.}\ \bibnamefont {de~Pablo}},\ and\
  \bibinfo {author} {\bibfnamefont {M.~L.}\ \bibnamefont {Gardel}},\ }\bibfield
   {title} {\bibinfo {title} {Tunable structure and dynamics of active liquid
  crystals},\ }\bibfield  {journal} {\bibinfo  {journal} {Science Advances}\
  }\textbf {\bibinfo {volume} {4}},\ \href
  {https://doi.org/10.1126/sciadv.aat7779} {10.1126/sciadv.aat7779} (\bibinfo
  {year} {2018})\BibitemShut {NoStop}%
\bibitem [{\citenamefont {F{\"u}rthauer}\ \emph {et~al.}(2019)\citenamefont
  {F{\"u}rthauer}, \citenamefont {Lemma}, \citenamefont {Foster}, \citenamefont
  {Ems-McClung}, \citenamefont {Yu}, \citenamefont {Walczak}, \citenamefont
  {Dogic}, \citenamefont {Needleman},\ and\ \citenamefont
  {Shelley}}]{Fuerthauer2019}%
  \BibitemOpen
  \bibfield  {author} {\bibinfo {author} {\bibfnamefont {S.}~\bibnamefont
  {F{\"u}rthauer}}, \bibinfo {author} {\bibfnamefont {B.}~\bibnamefont
  {Lemma}}, \bibinfo {author} {\bibfnamefont {P.~J.}\ \bibnamefont {Foster}},
  \bibinfo {author} {\bibfnamefont {S.~C.}\ \bibnamefont {Ems-McClung}},
  \bibinfo {author} {\bibfnamefont {C.-H.}\ \bibnamefont {Yu}}, \bibinfo
  {author} {\bibfnamefont {C.~E.}\ \bibnamefont {Walczak}}, \bibinfo {author}
  {\bibfnamefont {Z.}~\bibnamefont {Dogic}}, \bibinfo {author} {\bibfnamefont
  {D.~J.}\ \bibnamefont {Needleman}},\ and\ \bibinfo {author} {\bibfnamefont
  {M.~J.}\ \bibnamefont {Shelley}},\ }\bibfield  {title} {\bibinfo {title}
  {Self-straining of actively crosslinked microtubule networks},\ }\href
  {https://doi.org/10.1038/s41567-019-0642-1} {\bibfield  {journal} {\bibinfo
  {journal} {Nature Physics}\ }\textbf {\bibinfo {volume} {15}},\ \bibinfo
  {pages} {1295} (\bibinfo {year} {2019})}\BibitemShut {NoStop}%
\bibitem [{\citenamefont {Ludwig}\ \emph {et~al.}(2020)\citenamefont {Ludwig},
  \citenamefont {Weirch}, \citenamefont {Alster}, \citenamefont {Witten},
  \citenamefont {Gardel}, \citenamefont {Dasbiswas},\ and\ \citenamefont
  {Vaikuntanathan}}]{Ludwig20}%
  \BibitemOpen
  \bibfield  {author} {\bibinfo {author} {\bibfnamefont {N.~B.}\ \bibnamefont
  {Ludwig}}, \bibinfo {author} {\bibfnamefont {K.~L.}\ \bibnamefont {Weirch}},
  \bibinfo {author} {\bibfnamefont {E.}~\bibnamefont {Alster}}, \bibinfo
  {author} {\bibfnamefont {T.~A.}\ \bibnamefont {Witten}}, \bibinfo {author}
  {\bibfnamefont {M.~L.}\ \bibnamefont {Gardel}}, \bibinfo {author}
  {\bibfnamefont {K.}~\bibnamefont {Dasbiswas}},\ and\ \bibinfo {author}
  {\bibfnamefont {S.}~\bibnamefont {Vaikuntanathan}},\ }\bibfield  {title}
  {\bibinfo {title} {Nucleation and shape dynamics of model nematic tactoids
  around adhesive colloids},\ }\href {https://doi.org/10.1063/1.5141997}
  {\bibfield  {journal} {\bibinfo  {journal} {The Journal of Chemical Physics}\
  }\textbf {\bibinfo {volume} {152}},\ \bibinfo {pages} {084901} (\bibinfo
  {year} {2020})},\ \Eprint
  {https://arxiv.org/abs/https://doi.org/10.1063/1.5141997}
  {https://doi.org/10.1063/1.5141997} \BibitemShut {NoStop}%
\bibitem [{\citenamefont {Toner}\ and\ \citenamefont {Tu}(1995)}]{Toner1995}%
  \BibitemOpen
  \bibfield  {author} {\bibinfo {author} {\bibfnamefont {J.}~\bibnamefont
  {Toner}}\ and\ \bibinfo {author} {\bibfnamefont {Y.}~\bibnamefont {Tu}},\
  }\bibfield  {title} {\bibinfo {title} {Long-range order in a two-dimensional
  dynamical $\mathrm{XY}$ model: How birds fly together},\ }\href
  {https://doi.org/10.1103/PhysRevLett.75.4326} {\bibfield  {journal} {\bibinfo
   {journal} {Phys. Rev. Lett.}\ }\textbf {\bibinfo {volume} {75}},\ \bibinfo
  {pages} {4326} (\bibinfo {year} {1995})}\BibitemShut {NoStop}%
\bibitem [{\citenamefont {Husain}\ and\ \citenamefont
  {Rao}(2017)}]{Husain2017}%
  \BibitemOpen
  \bibfield  {author} {\bibinfo {author} {\bibfnamefont {K.}~\bibnamefont
  {Husain}}\ and\ \bibinfo {author} {\bibfnamefont {M.}~\bibnamefont {Rao}},\
  }\bibfield  {title} {\bibinfo {title} {Emergent structures in an active polar
  fluid: Dynamics of shape, scattering, and merger},\ }\href
  {https://doi.org/10.1103/PhysRevLett.118.078104} {\bibfield  {journal}
  {\bibinfo  {journal} {Phys. Rev. Lett.}\ }\textbf {\bibinfo {volume} {118}},\
  \bibinfo {pages} {078104} (\bibinfo {year} {2017})}\BibitemShut {NoStop}%
\bibitem [{\citenamefont {Liebchen}\ and\ \citenamefont
  {L\"owen}(2018)}]{liebchen2018synthetic}%
  \BibitemOpen
  \bibfield  {author} {\bibinfo {author} {\bibfnamefont {B.}~\bibnamefont
  {Liebchen}}\ and\ \bibinfo {author} {\bibfnamefont {H.}~\bibnamefont
  {L\"owen}},\ }\bibfield  {title} {\bibinfo {title} {Synthetic chemotaxis and
  collective behavior in active matter},\ }\href@noop {} {\bibfield  {journal}
  {\bibinfo  {journal} {Accounts of chemical research}\ }\textbf {\bibinfo
  {volume} {51}},\ \bibinfo {pages} {2982} (\bibinfo {year}
  {2018})}\BibitemShut {NoStop}%
\bibitem [{\citenamefont {Kung}\ \emph {et~al.}(2006)\citenamefont {Kung},
  \citenamefont {Cristina~Marchetti},\ and\ \citenamefont {Saunders}}]{Kung06}%
  \BibitemOpen
  \bibfield  {author} {\bibinfo {author} {\bibfnamefont {W.}~\bibnamefont
  {Kung}}, \bibinfo {author} {\bibfnamefont {M.}~\bibnamefont
  {Cristina~Marchetti}},\ and\ \bibinfo {author} {\bibfnamefont
  {K.}~\bibnamefont {Saunders}},\ }\bibfield  {title} {\bibinfo {title}
  {Hydrodynamics of polar liquid crystals},\ }\href
  {https://doi.org/10.1103/PhysRevE.73.031708} {\bibfield  {journal} {\bibinfo
  {journal} {Phys. Rev. E}\ }\textbf {\bibinfo {volume} {73}},\ \bibinfo
  {pages} {031708} (\bibinfo {year} {2006})}\BibitemShut {NoStop}%
\bibitem [{\citenamefont {Everts}\ \emph {et~al.}(2016)\citenamefont {Everts},
  \citenamefont {Punter}, \citenamefont {Samin}, \citenamefont {van~der
  Schoot},\ and\ \citenamefont {van Roij}}]{Everts2016}%
  \BibitemOpen
  \bibfield  {author} {\bibinfo {author} {\bibfnamefont {J.~C.}\ \bibnamefont
  {Everts}}, \bibinfo {author} {\bibfnamefont {M.~T. J. J.~M.}\ \bibnamefont
  {Punter}}, \bibinfo {author} {\bibfnamefont {S.}~\bibnamefont {Samin}},
  \bibinfo {author} {\bibfnamefont {P.}~\bibnamefont {van~der Schoot}},\ and\
  \bibinfo {author} {\bibfnamefont {R.}~\bibnamefont {van Roij}},\ }\bibfield
  {title} {\bibinfo {title} {A landau-de gennes theory for hard colloidal rods:
  Defects and tactoids},\ }\href {https://doi.org/10.1063/1.4948785} {\bibfield
   {journal} {\bibinfo  {journal} {The Journal of Chemical Physics}\ }\textbf
  {\bibinfo {volume} {144}},\ \bibinfo {pages} {194901} (\bibinfo {year}
  {2016})},\ \Eprint {https://arxiv.org/abs/https://doi.org/10.1063/1.4948785}
  {https://doi.org/10.1063/1.4948785} \BibitemShut {NoStop}%
\bibitem [{\citenamefont {Safran}(2018)}]{safran2018statistical}%
  \BibitemOpen
  \bibfield  {author} {\bibinfo {author} {\bibfnamefont {S.}~\bibnamefont
  {Safran}},\ }\href@noop {} {\emph {\bibinfo {title} {Statistical
  thermodynamics of surfaces, interfaces, and membranes}}}\ (\bibinfo
  {publisher} {CRC Press},\ \bibinfo {year} {2018})\BibitemShut {NoStop}%
\bibitem [{\citenamefont {Pettey}\ and\ \citenamefont
  {Lubensky}(1999)}]{Pettey99}%
  \BibitemOpen
  \bibfield  {author} {\bibinfo {author} {\bibfnamefont {D.}~\bibnamefont
  {Pettey}}\ and\ \bibinfo {author} {\bibfnamefont {T.~C.}\ \bibnamefont
  {Lubensky}},\ }\bibfield  {title} {\bibinfo {title} {Stability of texture and
  shape of circular domains of langmuir monolayers},\ }\href
  {https://doi.org/10.1103/PhysRevE.59.1834} {\bibfield  {journal} {\bibinfo
  {journal} {Phys. Rev. E}\ }\textbf {\bibinfo {volume} {59}},\ \bibinfo
  {pages} {1834} (\bibinfo {year} {1999})}\BibitemShut {NoStop}%
\bibitem [{\citenamefont {Ganem}\ \emph {et~al.}(2009)\citenamefont {Ganem},
  \citenamefont {Godinho},\ and\ \citenamefont {Pellman}}]{Ganem2009}%
  \BibitemOpen
  \bibfield  {author} {\bibinfo {author} {\bibfnamefont {N.~J.}\ \bibnamefont
  {Ganem}}, \bibinfo {author} {\bibfnamefont {S.~A.}\ \bibnamefont {Godinho}},\
  and\ \bibinfo {author} {\bibfnamefont {D.}~\bibnamefont {Pellman}},\
  }\bibfield  {title} {\bibinfo {title} {A mechanism linking extra centrosomes
  to chromosomal instability},\ }\href {https://doi.org/10.1038/nature08136}
  {\bibfield  {journal} {\bibinfo  {journal} {Nature}\ }\textbf {\bibinfo
  {volume} {460}},\ \bibinfo {pages} {278} (\bibinfo {year}
  {2009})}\BibitemShut {NoStop}%
\bibitem [{\citenamefont {Metselaar}\ \emph {et~al.}(2019)\citenamefont
  {Metselaar}, \citenamefont {Yeomans},\ and\ \citenamefont
  {Doostmohammadi}}]{Metselaar2019}%
  \BibitemOpen
  \bibfield  {author} {\bibinfo {author} {\bibfnamefont {L.}~\bibnamefont
  {Metselaar}}, \bibinfo {author} {\bibfnamefont {J.~M.}\ \bibnamefont
  {Yeomans}},\ and\ \bibinfo {author} {\bibfnamefont {A.}~\bibnamefont
  {Doostmohammadi}},\ }\bibfield  {title} {\bibinfo {title} {Topology and
  morphology of self-deforming active shells},\ }\href
  {https://doi.org/10.1103/PhysRevLett.123.208001} {\bibfield  {journal}
  {\bibinfo  {journal} {Phys. Rev. Lett.}\ }\textbf {\bibinfo {volume} {123}},\
  \bibinfo {pages} {208001} (\bibinfo {year} {2019})}\BibitemShut {NoStop}%
\bibitem [{\citenamefont {Oriola}\ \emph {et~al.}(2020)\citenamefont {Oriola},
  \citenamefont {J{\"u}licher},\ and\ \citenamefont
  {Brugu{\'e}s}}]{Oriola2020}%
  \BibitemOpen
  \bibfield  {author} {\bibinfo {author} {\bibfnamefont {D.}~\bibnamefont
  {Oriola}}, \bibinfo {author} {\bibfnamefont {F.}~\bibnamefont
  {J{\"u}licher}},\ and\ \bibinfo {author} {\bibfnamefont {J.}~\bibnamefont
  {Brugu{\'e}s}},\ }\bibfield  {title} {\bibinfo {title} {Active forces shape
  the metaphase spindle through a mechanical instability},\ }\href
  {https://doi.org/10.1073/pnas.2002446117} {\bibfield  {journal} {\bibinfo
  {journal} {Proceedings of the National Academy of Sciences}\ }\textbf
  {\bibinfo {volume} {117}},\ \bibinfo {pages} {16154} (\bibinfo {year}
  {2020})},\ \Eprint
  {https://arxiv.org/abs/https://www.pnas.org/content/117/28/16154.full.pdf}
  {https://www.pnas.org/content/117/28/16154.full.pdf} \BibitemShut {NoStop}%
\bibitem [{\citenamefont {Roostalu}\ \emph {et~al.}(2018)\citenamefont
  {Roostalu}, \citenamefont {Rickman}, \citenamefont {Thomas}, \citenamefont
  {Nédélec},\ and\ \citenamefont {Surrey}}]{Roostalu2018}%
  \BibitemOpen
  \bibfield  {author} {\bibinfo {author} {\bibfnamefont {J.}~\bibnamefont
  {Roostalu}}, \bibinfo {author} {\bibfnamefont {J.}~\bibnamefont {Rickman}},
  \bibinfo {author} {\bibfnamefont {C.}~\bibnamefont {Thomas}}, \bibinfo
  {author} {\bibfnamefont {F.}~\bibnamefont {Nédélec}},\ and\ \bibinfo
  {author} {\bibfnamefont {T.}~\bibnamefont {Surrey}},\ }\bibfield  {title}
  {\bibinfo {title} {Determinants of polar versus nematic organization in
  networks of dynamic microtubules and mitotic motors},\ }\href
  {https://doi.org/https://doi.org/10.1016/j.cell.2018.09.029} {\bibfield
  {journal} {\bibinfo  {journal} {Cell}\ }\textbf {\bibinfo {volume} {175}},\
  \bibinfo {pages} {796 } (\bibinfo {year} {2018})}\BibitemShut {NoStop}%
\bibitem [{\citenamefont {Leoni}\ \emph {et~al.}(2017)\citenamefont {Leoni},
  \citenamefont {Manyuhina}, \citenamefont {Bowick},\ and\ \citenamefont
  {Marchetti}}]{Leoni2017}%
  \BibitemOpen
  \bibfield  {author} {\bibinfo {author} {\bibfnamefont {M.}~\bibnamefont
  {Leoni}}, \bibinfo {author} {\bibfnamefont {O.~V.}\ \bibnamefont
  {Manyuhina}}, \bibinfo {author} {\bibfnamefont {M.~J.}\ \bibnamefont
  {Bowick}},\ and\ \bibinfo {author} {\bibfnamefont {M.~C.}\ \bibnamefont
  {Marchetti}},\ }\bibfield  {title} {\bibinfo {title} {Defect driven shapes in
  nematic droplets: analogies with cell division},\ }\href
  {https://doi.org/10.1039/C6SM02584F} {\bibfield  {journal} {\bibinfo
  {journal} {Soft Matter}\ }\textbf {\bibinfo {volume} {13}},\ \bibinfo {pages}
  {1257} (\bibinfo {year} {2017})}\BibitemShut {NoStop}%
\bibitem [{\citenamefont {Holmes}(2019)}]{Holmes2019}%
  \BibitemOpen
  \bibfield  {author} {\bibinfo {author} {\bibfnamefont {D.~P.}\ \bibnamefont
  {Holmes}},\ }\bibfield  {title} {\bibinfo {title} {Elasticity and stability
  of shape-shifting structures},\ }\href@noop {} {\bibfield  {journal}
  {\bibinfo  {journal} {Current opinion in colloid \& interface science}\
  }\textbf {\bibinfo {volume} {40}},\ \bibinfo {pages} {118} (\bibinfo {year}
  {2019})},\ \bibinfo {note}
  {\url{http://doi.org/10.1016/j.cocis.2019.02.008}}\BibitemShut {NoStop}%
\bibitem [{\citenamefont {Liouville}(1853)}]{liouville1853}%
  \BibitemOpen
  \bibfield  {author} {\bibinfo {author} {\bibfnamefont {J.}~\bibnamefont
  {Liouville}},\ }\bibfield  {title} {\bibinfo {title} {Sur l’équation aux
  différences partielles ${\displaystyle {\frac {d^{2}\log \lambda }{dudv}}\pm
  {\frac {\lambda }{2a^{2}}}=0} {\displaystyle {\frac {d^{2}\log \lambda
  }{dudv}}\pm {\frac {\lambda }{2a^{2}}}=0}$.},\ }\href@noop {} {\bibfield
  {journal} {\bibinfo  {journal} {Journal de mathématiques pures et
  appliquées}\ ,\ \bibinfo {pages} {71}} (\bibinfo {year} {1853})}\BibitemShut
  {NoStop}%
\bibitem [{\citenamefont {Bratu}(1914)}]{bratu1914}%
  \BibitemOpen
  \bibfield  {author} {\bibinfo {author} {\bibfnamefont {G.}~\bibnamefont
  {Bratu}},\ }\bibfield  {title} {\bibinfo {title} {Sur les équations
  intégrales non linéaires.},\ }\href@noop {} {\bibfield  {journal} {\bibinfo
   {journal} {Bulletin de la Société Mathématique de France}\ }\textbf
  {\bibinfo {volume} {42}},\ \bibinfo {pages} {113–142} (\bibinfo {year}
  {1914})}\BibitemShut {NoStop}%
\bibitem [{\citenamefont {Gelfand}(1963)}]{gelfand1963}%
  \BibitemOpen
  \bibfield  {author} {\bibinfo {author} {\bibfnamefont {I.~M.}\ \bibnamefont
  {Gelfand}},\ }\bibfield  {title} {\bibinfo {title} {Some problems in the
  theory of quasilinear equations.},\ }\href@noop {} {\bibfield  {journal}
  {\bibinfo  {journal} {Amer. Math. Soc. Transl}\ }\textbf {\bibinfo {volume}
  {29.2}},\ \bibinfo {pages} {295–381} (\bibinfo {year} {1963})}\BibitemShut
  {NoStop}%
\bibitem [{\citenamefont {Walker}(1915)}]{walker1915}%
  \BibitemOpen
  \bibfield  {author} {\bibinfo {author} {\bibfnamefont {G.~W.}\ \bibnamefont
  {Walker}},\ }\bibfield  {title} {\bibinfo {title} {Some problems illustrating
  the forms of nebulae.},\ }\href@noop {} {\bibfield  {journal} {\bibinfo
  {journal} {Proceedings of the Royal Society of London. Series A, Containing
  Papers of a Mathematical and Physical Character}\ }\textbf {\bibinfo {volume}
  {91.631}},\ \bibinfo {pages} {410} (\bibinfo {year} {1915})}\BibitemShut
  {NoStop}%
\end{thebibliography}%

\clearpage

\onecolumngrid

\renewcommand{\thefigure}{S\arabic{figure}}
\renewcommand{\theequation}{S\arabic{equation}}
\setcounter{figure}{0}    
\setcounter{equation}{0}    

\begin{center}
    {\Large Supplementary Information}
\end{center}

%\section{Appendixes}
\section{Derivation and solution of  one dimensional model}

The 1D model for the dynamics of filament density, filament polarization and motor density given in the main text follow from the  conservation equations:
\begin{align}
\partial_t n_{+}&=D \partial_{x}^{2} n_{+} - \zeta \partial_{x}\left(n_{+} m\right),
\label{eq:nplus}
\\ 
\partial_t n_{-}&=D \partial_{x}^{2} n_{-} + \zeta\partial_{x}\left(n_{-} m\right),
\label{eq:nminus}
\\
\partial_t m &=D_m \partial_x^2 m + v_0 \partial_x m(n_+ -n_- ),
\label{eq:cnpnm}
\end{align}

Our one dimensional model (Eqs.~\eqref{eq:rho1d}-\eqref{eq:motor1d}) is simplified by assuming incompressibility, i.e. the density is $\rho = \rho_0$. 
In the steady state Eqs.~\eqref{eq:rho1d}-\eqref{eq:motor1d} then reduce to
\begin{align}
\partial_{t} p&= D\partial_{x}^{2} p  - \zeta \rho_0 \partial_{x} m,
\label{eq:p1dincomp}
\\
\partial_t m &= D_m\partial_x^2 m + v_0 \partial_x (m p).
\label{eq:motor1dcomp}
\end{align}
The Eqs.~\eqref{eq:p1dincomp}-\eqref{eq:motor1dcomp} are solved with ``natural boundary''conditions that the motor density and polarization as well as their corresponding fluxes decay to zero at $x=\pm \infty$  to yield:
\begin{align}
&p(x) = \mathrm{tanh} \left(  x/\xi \right),
\label{eq:1dpsolution}\\
&m(x) = m_0 \mathrm{sech}^2 \left( x/\xi \right),
\label{eq:1dmsolution}
\end{align}
where $\xi= \frac{2D_m}{v_0}$ is the typical aster length scale and $m_0 = \frac{D \xi}{\zeta \rho_0}$ is the aster strength. 
Figure~\ref{fig:sketch1d}(ii-iii) shows the solutions Eqs.~\eqref{eq:1dpsolution}-\eqref{eq:1dmsolution}, which show the one dimensional equivalent of an  out pointing aster. The motors gather at the center of the aster as expected. Note that in 1D, the flux at steady state is a constant throughout the system, and that natural boundary conditions require this flux to vanish. In other words the diffusive flux of the motors and filaments is balanced by the corresponding active flux that advects them, at steady state. This competition sets the width of the distribution of motors at the ``aster'': $\xi \sim \frac{D_m}{v_0}$. In a confined droplet geometry, and at higher dimensions, solutions with nonzero fluxes are possible and seen in our numeric solutions.
%\begin{figure}
%    \centering
%    \includegraphics[width=0.49\textwidth]{Figures/1d_aster.pdf}
%    \caption{Caption: Solution of one dimensional incompressible model. Red dashed line shows the polarization Eq.~\eqref{eq:1dpsolution} and red solid line shows the motor concentration Eq.~\eqref{eq:1dmsolution}. }
    \label{fig:1daster}
%\end{figure}
 %\section{Phenomenological derivation of 2D  model}

\section{1D numerics}
 \begin{figure*}
     \centering
     \includegraphics[width=1.0\textwidth]{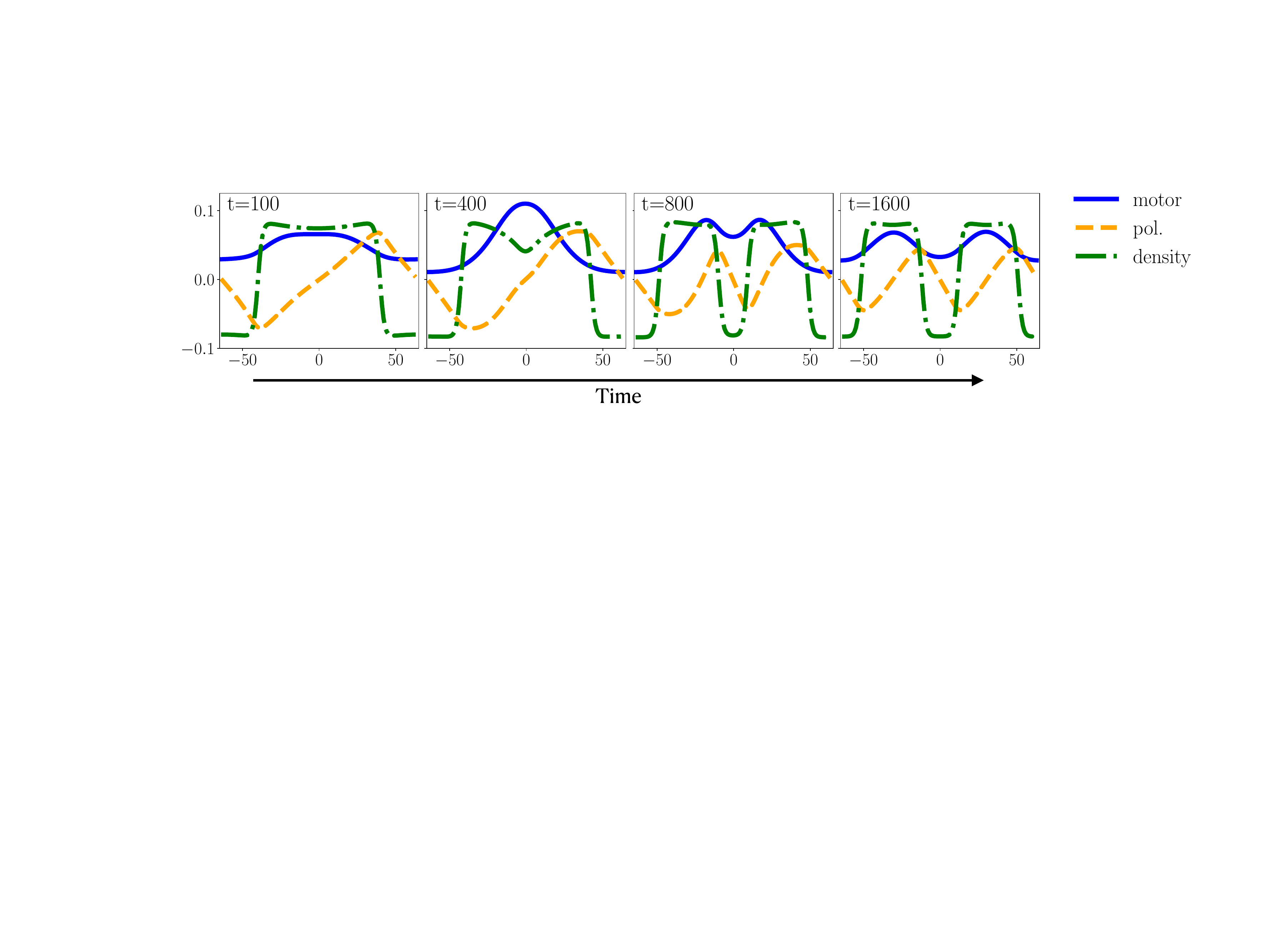}
     \caption{Division of a one dimensional droplet. Each plot shows a time snapshot with the time shown in the upper left corner. Blue solid lines show the motor concentration, orange dashed lines show the polarization and green dashed dotted lines show the density.}
     \label{fig:1ddropletdivision}
 \end{figure*}
 
 \begin{figure}
     \centering
     \includegraphics[width=0.5\textwidth]{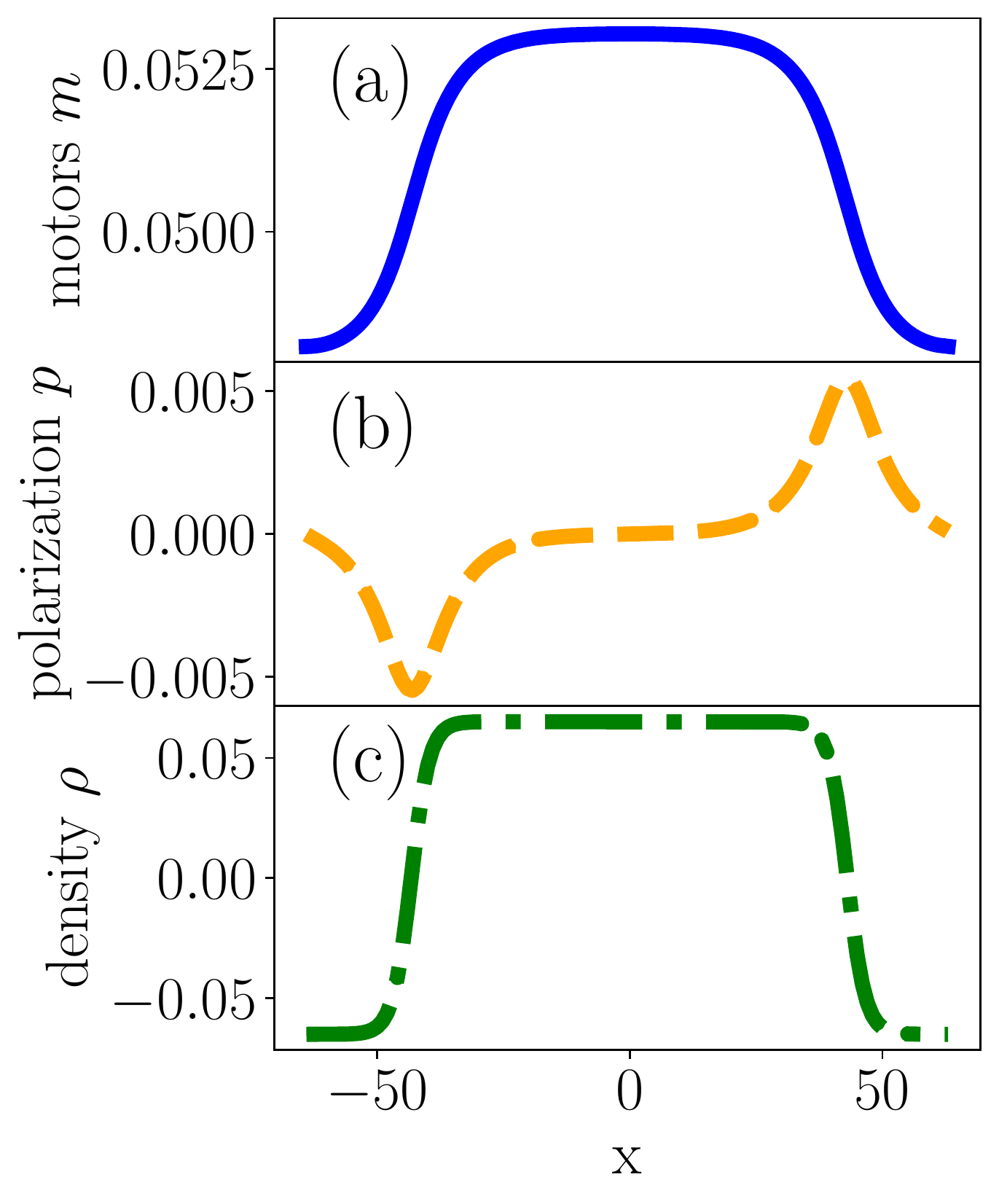}
     \caption{Single droplet with centering aster in one dimensional model. (a) Motor concentration. (b) Polarization o filaments. (c) Density of filaments. }
     \label{fig:1dmodelSingleDroplet}
 \end{figure}
 
To test the minimal 1D model in a finite domain, we numerically integrated the following equations
\begin{align}
\partial_{t} \rho&=D\partial_x^{2} \rho - \zeta\rho_0 \partial_{x}(p m) +  \Gamma_{\rho} \partial_x^{2} 
\frac{\delta F_{\mathrm{1d} }}{\delta \rho} ,
\label{eq:rho1dnum}
\\ 
\partial_{t} p&= D\partial_{x}^{2} p  - \zeta/\rho_0 \partial_{x}(\rho m) - \Gamma_{p} \frac{\delta F_{\mathrm{1d}}}{\delta p},
\label{eq:p1dnum}
\\
\partial_t m &= D_m\partial_x^2 m + v_0 \partial_x (m p) .
\label{eq:motor1dnum}
\end{align}
 Here, we included the following free energy
\begin{align}
F_{\mathrm{1d}}=&-\frac{\nu_{2}}{2} \rho^{2}%-\frac{\nu_{3}}{3} \psi^{3}
    +\frac{\nu_4}{4} \rho^{4}+\frac{B}{2}( \partial_x \rho)^{2}\\
    &+\alpha_p p^2/2 +\beta_p p^4/2,
    \label{eq:freeE1dnum}
\end{align}
 which has a phase field for the density $\rho$ such that a one dimensional droplet is formed and $F_{\mathrm{1d}}$ has decaying terms for the polarization $p$. 
 
 We integrate Eq.\eqref{eq:rho1dnum}-\eqref{eq:motor1dnum} numerically using a a finite difference method with a forth order Runge-Kutta time discretization and periodic boundary conditions. The simulation parameters can be found in table~\ref{tab:simparameters_onedim}.
\begin{table}
\begin{tabular}{|l|l|}
\hline
\textbf{Parameter} & \textbf{Value}
\\\hline
$\Delta x$ & $1.0$
\\\hline
grid & $128$
\\\hline
$\delta t$ & $0.001$
\\\hline
$D_m$ & $2$
\\\hline
$v_0$& $2$
\\\hline
$\zeta$& $10.0$
\\\hline
$\kappa_p$ & $4.0$
\\\hline
$\alpha_p$ & $0.1$
\\\hline
$\beta_p$ & $1.0$
\\\hline
$\Gamma_p$ & $1.0$
\\\hline
$\nu_2$ & $0.01$
\\\hline
$\nu_4$ & $1.4$ 
\\\hline
$B$ & $0.04$
\\\hline
$\Gamma_{\psi}$ & $1.0$
\\\hline
$R_0$ & $35$
\\\hline
$c_0$ & $0.1$
\\\hline
\end{tabular}
\caption{Parameter values used in one dimensional model.}
\label{tab:simparameters_onedim}
\end{table}
 
 Figure~\ref{fig:1dmodelSingleDroplet} show the motor concentration (Fig.~\ref{fig:1dmodelSingleDroplet}(a)), polarization (Fig.~\ref{fig:1dmodelSingleDroplet}(b)) and density (Fig.~\ref{fig:1dmodelSingleDroplet}(c)) for a simulation at low activity $\zeta=1.5$. We find the emergence of a single droplet with a centering aster.
 
 Going to higher activity $\zeta=3.5$, we find that the droplet can divide by means of the active motion of filaments. Figure~\ref{fig:1ddropletdivision} shows a time series of the division of a droplet. Here, the blue solid line show the motor concentration, the dashed orange line the polarization, and the dashed dotted green line the density.

%\section{High friction approximation of fluid on surface}
%The hydrodynamic interactions at low Reynolds  number are governed by the Stokes equations
%\begin{align}
%\eta\nabla^2\bm{u}
%=\nabla p + \Gamma_s \bm{u} -\nabla \cdot \mathbf{\sigma}_a, \nonumber \\  %\nabla\cdot\bm{u}=0.
%\label{eq:Stokes}
%\end{align}
%Here, $\bm{u}$ is the velocity field of the surrounding fluid, $\eta$ is the viscosity, $p$ is the fluids' pressure, $\Gamma_s$ is the friction with the surface and $\sigma_a$ are the active stresses. 
%Given that the flow is dominated by the active motion and the friction with the surface, we can approximate the flow by
%\begin{align}
%\Gamma_s \bm{u} \approx \nabla \cdot \mathbf{\sigma}_a.
%\label{eq:Stokesapprox}
%\end{align}
%In the main text we work with in this approximation and neglect all long ranged hydrodynamic flows.
%integrate out velocity from momentum balance equation on substrate to derive dry model 

\section{Simulation method}

We solve Eq.~\eqref{eq:c_equation}-\eqref{eq:Q_equation} numerically using  a finite difference scheme with a fourth order Runge-Kutta time integration  on a $74\times 74$ grid. We use a timestep of $\delta t= 0.01$ and the various parameters are given in Table~\ref{tab:simparameters}, while we vary surface tension, $\gamma$, and the strength of motor-induced density flux and molecular torque: $\zeta$ and $\zeta_{p}$. We initialize the system with a circular droplet of radius $R_0$.

The parameters used in the simulations in the main text are given in Table~\ref{tab:exampleparamters} and Table~\ref{tab:simparameters} .

\begin{table}
\begin{tabular}{|l|l|l|l|l|}
\hline
\textbf{Parameter} & \textbf{Fig.~\ref{fig:kinetic_pathway}iiia} %formerly 2a}
& \textbf{Fig.2(c)} & \textbf{Fig.2(h)} 
& \textbf{Fig.2(j)}
\\\hline
$\Delta x \ \ \Delta y$ & $1.0 \ \ 1.0$
& $1.0 \ \ 1.0$ & $1.0 \ \ 1.0$ & $1.0 \ \ 1.0$
\\\hline
grid & $100\times 100$ & $100\times 100$ & $100\times 100$ & $100\times 100$
\\\hline
$D_m$ & $0.4$ & $0.4$ & $0.4$ & $0.4$
\\\hline
$v_0$& $0.15$ & $0.15$ & $0.15$ & $0.15$
\\\hline
$k_{\mathrm{on}}$ &$0.00005$ &$0.00015$ &$0.00015$ &$0.00015$
\\\hline
$k_{\mathrm{off}}$ & $0.0005$ & $0.001$  & $0.001$  & $0.001$
\\\hline
$\nu_2$ & $0.4$ & $0.4$ & $0.4$ & $0.4$ 
\\\hline
$\nu_4$ & $2.0$ & $2.0$& $2.0$& $2.0$
\\\hline
$\Gamma_{\psi}$ &$1.0$ &$1.0$&$1.0$&$1.0$
\\\hline
$R_0$ & $23.0$ & $23.0$& $23.0$& $23.0$
\\\hline
$\kappa_p$ & $0.1$& $0.1$& $0.1$& $0.1$
\\\hline
$\alpha_p$ & $0.1$& $0.1$& $0.1$& $0.1$
\\\hline
$\beta_p$ & $10.0$& $10.0$& $10.0$& $10.0$
\\\hline
$\Gamma_p$ & $1.0$& $1.0$& $1.0$& $1.0$
\\\hline
$\zeta_0$ & $0.01$& $0.01$& $0.01$& $0.01$
\\\hline
$\kappa_Q$ & $0.55$& $0.4$& $0.4$& $0.4$
\\\hline
$\alpha_Q$ &$ 0.01$&$ 0.01$&$ 0.01$&$ 0.01$
\\\hline
$\beta_Q$ & $2.0$& $2.0$& $2.0$& $2.0$
\\\hline
$\Gamma_Q$ & $1.0$& $1.0$& $1.0$& $1.0$
\\\hline
$C$ & $1.0$& $1.0$& $1.0$& $1.0$
\\\hline
$A$ & $0.6$& $0.8$& $0.8$& $0.8$
\\\hline
$\Omega$ & $0.3$& $0.3$& $0.3$& $0.3$
\\\hline
$B$ & $1.8$ & $2.8$ & $2.8$ &  $2.8$
\\\hline
$\zeta$ & $4.0$ & $1.0$ & $0.8$ &  $1.0$
\\\hline
$\zeta_p$ & $20.0$ & $40.0$ & $5.0$ & $20.0$
\\\hline
\end{tabular}
\caption{Parameter values used for the example droplet shapes in Fig.~\ref{fig:kinetic_pathway}}
\label{tab:exampleparamters}
\end{table}

\newpage

\begin{table}
\begin{tabular}{|l|l|}
\hline
Fig.\ref{fig:combined_fig}(i),(vii)) & 
\\\hline
\textbf{Parameter} & \textbf{Value}
\\\hline
$\Delta x \ \ \Delta y$ & $1.0 \ \ 1.0$
\\\hline
grid & $74\times74$
\\\hline
$D_m$ & $0.4$
\\\hline
$v_0$& $0.15$
\\\hline
$k_{\mathrm{on}}$ &$0.00015$
\\\hline
$k_{\mathrm{off}}$ & $0.001$
\\\hline
$\nu_2$ & $0.4$ 
\\\hline
$\nu_4$ & $2.0$
\\\hline
$\Gamma_{\psi}$ &$1.0$
\\\hline
$R_0$ & $17.0$
\\\hline
$\kappa_p$ & $0.1$
\\\hline
$\alpha_p$ & $0.1$
\\\hline
$\beta_p$ & $10.0$
\\\hline
$\Gamma_p$ & $1.0$
\\\hline
$\zeta_0$ & $0.01$
\\\hline
$\kappa_Q$ & $0.3$
\\\hline
$\alpha_Q$ &$ 0.01$
\\\hline
$\beta_Q$ & $2.0$
\\\hline
$\Gamma_Q$ & $1.0$
\\\hline
$C$ & $1.0$
\\\hline
$A$ & $0.5$
\\\hline
$\Omega$ & $0.3$
\\\hline
\end{tabular}
\begin{tabular}{|l|l|}
\hline
Fig.\ref{fig:combined_fig}(vi)) & 
\\\hline
\textbf{Parameter} & \textbf{Value}
\\\hline
$\Delta x \ \ \Delta y$ & $1.0 \ \ 1.0$
\\\hline
grid & $74\times74$
\\\hline
$D_m$ & $0.4$
\\\hline
$v_0$& $0.15$
\\\hline
$k_{\mathrm{off}}$ & $0.001$
\\\hline
$\nu_2$ & $0.4$ 
\\\hline
$\nu_4$ & $2.0$
\\\hline
$\Gamma_{\psi}$ &$1.0$
\\\hline
$\kappa_p$ & $0.1$
\\\hline
$\alpha_p$ & $0.1$
\\\hline
$\beta_p$ & $10.0$
\\\hline
$\Gamma_p$ & $1.0$
\\\hline
$\zeta_0$ & $0.01$
\\\hline
$\kappa_Q$ & $0.3$
\\\hline
$\alpha_Q$ &$ 0.01$
\\\hline
$\beta_Q$ & $2.0$
\\\hline
$\Gamma_Q$ & $1.0$
\\\hline
$C$ & $1.0$
\\\hline
$A$ & $0.5$
\\\hline
$\Omega$ & $0.3$
\\\hline
$\zeta$ & $1.5$
\\\hline
$\zeta_p$ & $32.0$
\\\hline
\end{tabular}
\caption{Parameter values used in the simulations to compute the phase diagrams (Fig.\ref{fig:combined_fig}(i) and Fig.\ref{fig:combined_fig}(vii)) and to compute the critical motor concentration for three aster splitting as function of initial droplet size (Fig.\ref{fig:combined_fig}(vi)).}
\label{tab:simparameters}
\end{table}

%\begin{table}
%
%\caption{Parameter values used in the simulations }
%\label{tab:simparameters}
%\end{table}

%newpage

\section{Aster contribution to surface energy}

The formation of an aster is actively induced by the motor gradient term in Eq.~\ref{eq:p_equation}, which can be shown to arise from an effective free energy contribution, $ \zeta_{p}/\Gamma_{p} \int d^{2}x\, {\mathbf p} \cdot \nabla m$, to the polarization equation. Integrating by parts and setting the surface term to zero, this term can be seen to be equivalent to a spontaneous splay energy, 
$ - \zeta_{p}/\Gamma_{p} \int d^{2}x \, m \nabla \cdot {\mathbf p}$. 

For a uniform motor concentration, $m(x) = m_{0}$, this spontaneous splay term can be transformed into a surface term using the Green's theorem (which is also the 2D version of the divergence theorem) to give, $ - \zeta_{p} m_{0}/\Gamma_p \int dl \, \hat{{\mathbf n}}\cdot {\mathbf p}$, where the line element $dl$ is along the boundary of the droplet and $ \hat{n}$ is the unit normal to the droplet boundary. Since the ${\mathbf p}$ points outwards in the asters, we get $\hat{n}\cdot {\mathbf p} > 0$, which therefore contributes a \emph{negative line tension} to the droplet interfacial energy.  Comparing with the droplet line tension energy, $\gamma \int dl $, and assuming that polarization has a value $p_{0}$ at the droplet boundary,  we arrive at the condition, $\zeta_{p} p_{0} m_{0} > \Gamma_p \gamma$ for destabilization of droplet shape by the spontaneous splay induced by motor activity.

\section{Aster formation in bulk active polar model}
The gain more insight into the formation of multiple asters we study a simplified model for polar active fluids in bulk. 
We consider the following equations for motor concentration and polarization field,
\begin{align}
    &\partial_{t} m=D_m \Delta m+v_{0} \bm{\nabla} \cdot (m \bm{p}),
            \label{eq:simple_c_equation}
    \\     
    &\partial_{t} \bm{p}=-\zeta_p \bm{\nabla} m +\Gamma_{p} ( \kappa_p \Delta \bm{p} +\alpha_p \bm{p}  -\beta_p \bm{p} |\bm{p}|^2 ).
        \label{eq:simple_p_equation}
\end{align}
Here, we use $\Delta \equiv {\mathbf \nabla^{2}}$ is the Laplace-Beltrami operator In this simplified model, we consider a spontaneous long range order in the polarization instead of nematic order. We also neglect motor binding kinetics, and instead set the total number of motors to a constant average value, $m_{0}$,  which corresponds to the steady state of motor binding kinetics:  $m_{0} = k_{on} \psi_{0}/k_{off}$.  Further, we suppress the density kinetics by going to the incompressible limit and setting the compressibility $\zeta_0 =0$.

%which effectively models the ordering of filaments. 
Equations~\eqref{eq:simple_c_equation}-\eqref{eq:simple_p_equation} are integrated using a finite difference method with periodic boundary condition in a square box with side length $l$. For the time integration, we use a fourth order Runge-Kutta method with a timestep, $\delta t=0.01$.
We initialize the system with an aster in the polarization field $\bm{p}=  (\mathrm{cos} \phi, \mathrm{sin} \phi)^T$ and a uniform motor concentration, $m_0$. All simulation parameters are given in Table~\ref{tab:simparameters_bulk}.
\begin{table}
\begin{tabular}{|l|l|}
\hline
\textbf{Parameter} & \textbf{Value}
\\\hline
$\Delta x \ \ \Delta y$ & $1.0 \ \ 1.0$
\\\hline
$D_m$ & $0.5$
\\\hline
$v_0$& $0.35$
\\\hline
$\zeta_p$& $10.0$
\\\hline
$\kappa_p$ & $0.5$
\\\hline
$\alpha_p$ & $9.0$
\\\hline
$\beta_p$ & $10.0$
\\\hline
$\Gamma_p$ & $10.0$
\\\hline
\end{tabular}
\caption{Parameter values used in bulk simulations.}
\label{tab:simparameters_bulk}
\end{table}

\begin{figure}
    \centering
    \includegraphics[width=0.49\textwidth]{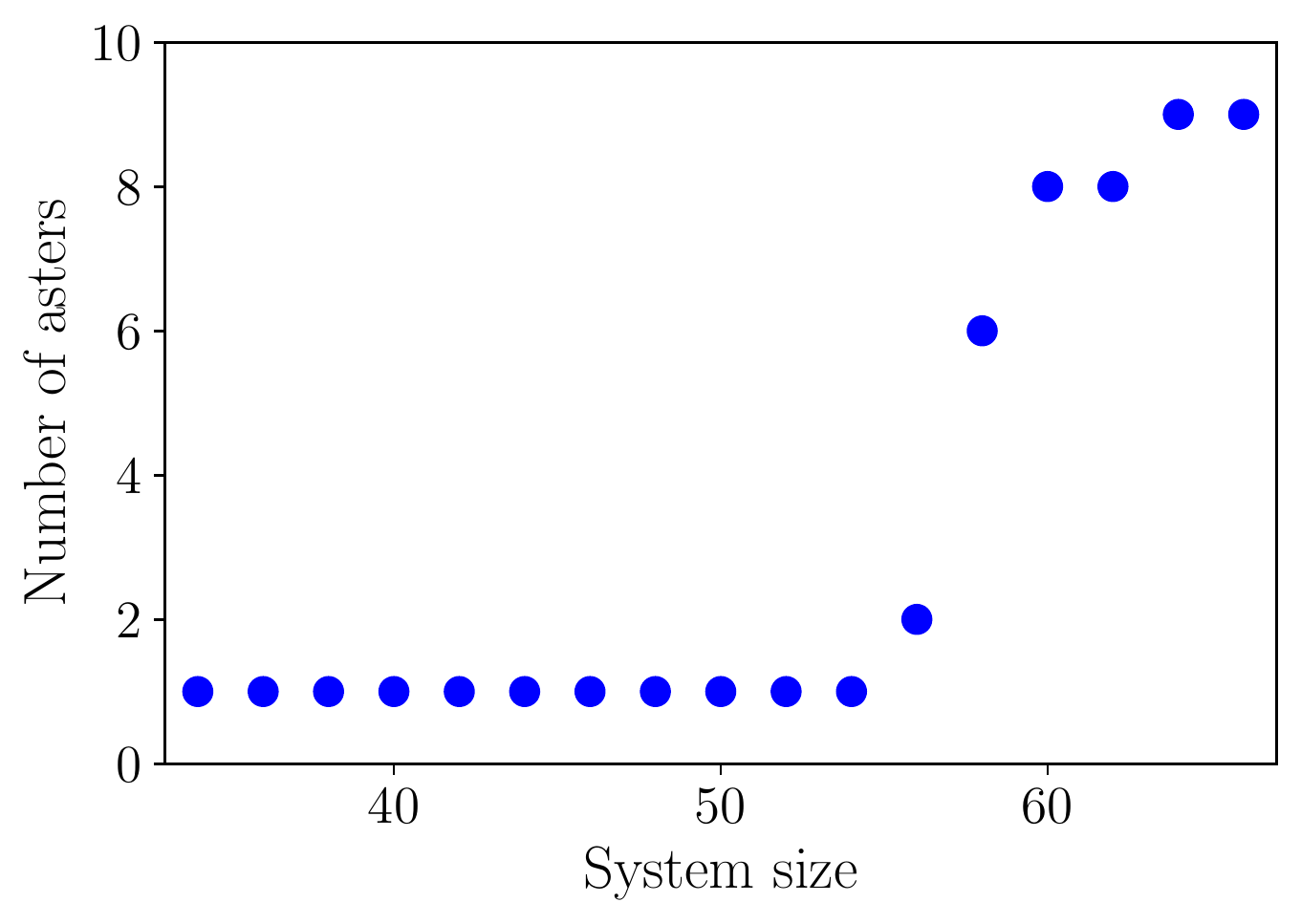}
    \caption{Number of aster as a function of system size.}
    \label{fig:bluk_aster_systemsize}
\end{figure}

First, we study the system size dependence by varying the box size $l$, and initial motor concentration $m_0=0.5$. In Fig.~\ref{fig:bluk_aster_systemsize} the number of asters as a function of system size $l$ is shown. The single aster state from the initial condition is destabilized above a critical system size $l_c = 54$ and we find a monotonic increase in number of asters. 
Linearzing Eq.\eqref{eq:simple_c_equation}-\eqref{eq:simple_p_equation} around the polarized state $\bm{p}= \bm{e}_y + \delta\bm{p}$, $m=m_0 +\delta m$ and transforming into Fourier space with wavevector $\bm{k}= (k_x,k_y)^T$ gives
\begin{align}
    &\partial_{t} \delta \tilde m= -D_m |\bm{k}|^2 \delta \tilde m + i v_{0}  m_0 \bm{k} \cdot \delta  \bm{\tilde p}
    + i v_{0}  i k_y \delta \tilde m,
            \label{eq:simple_c_equation_fourier}
    \\     
    &\partial_{t}  \delta \bm{\tilde p}=-\zeta \bm{k} \delta \tilde m - \Gamma_{p}  \kappa_p  |\bm{k}|^2\delta \bm{\tilde p},
        \label{eq:simple_p_equation_fourier}
\end{align}
where we neglected the Landau terms. This system has a maximal eigenvalue whose wavevector is given by
\begin{align}
    k_{\mathrm{max}} \approx  \frac{\sqrt{m_0 v_0 \zeta }}{D_m + \Gamma_p \kappa_p}.
    \label{eq:simple_max_wavevector}
\end{align}
The corresponding wavelength $L_{\mathrm{max}}=2 \pi/k_{\mathrm{max}}$ determines the grid spacing of asters in our simulation. Hence, we expect to find more than one aster when the system is larger then $2 L_{\mathrm{max}} = 52.2$ which agrees reasonably well with the system size we find in our simulations (see Fig.~\ref{fig:bluk_aster_systemsize}).

\begin{figure} [ht]
    \centering
    \includegraphics[width=0.49\textwidth]{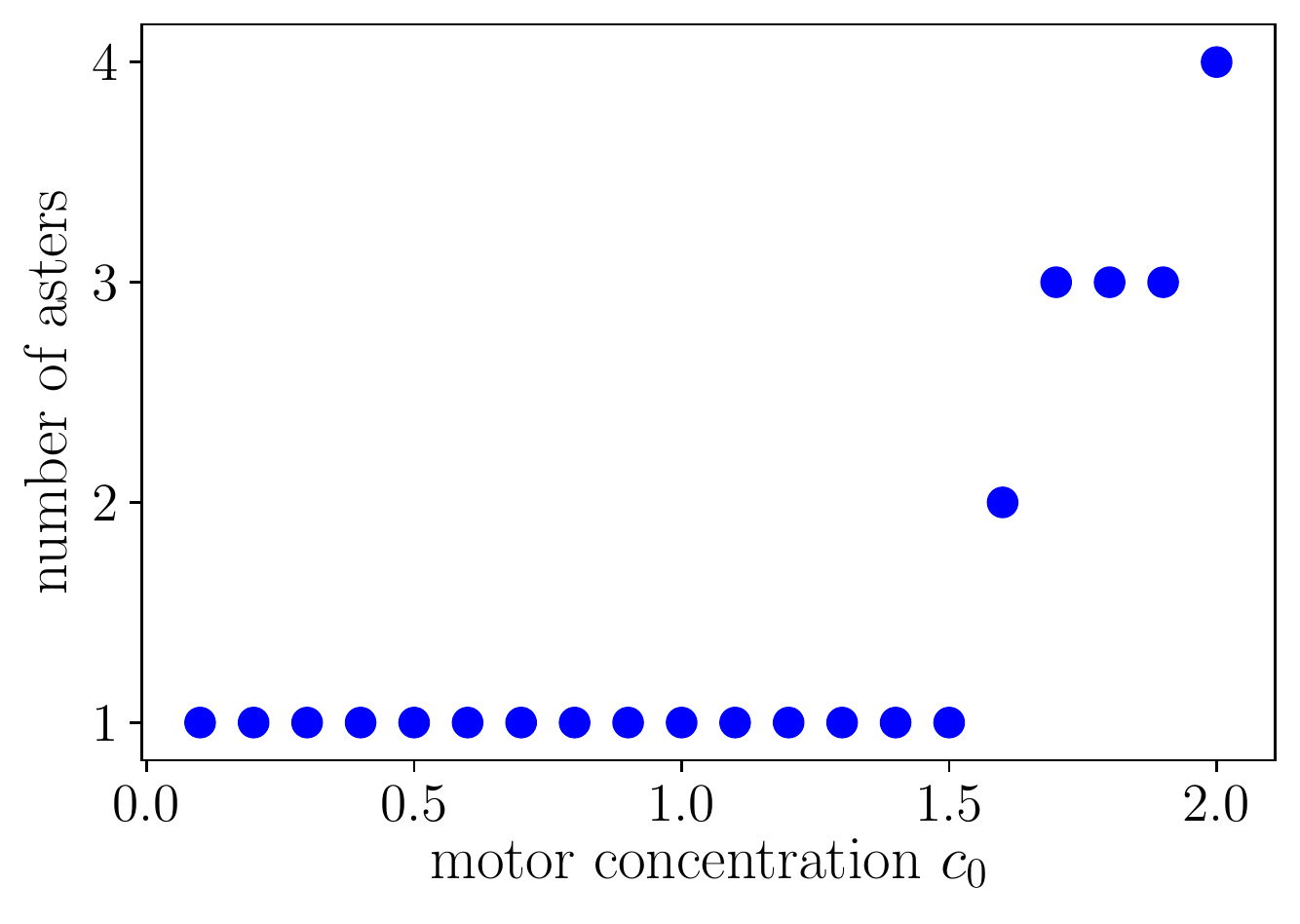}
    \caption{Number of asters as a function of bound motors $m_0$ in a bulk simulation. }
    \label{fig:naster_bulk}
\end{figure}

Second, we investigate the number of motors in the system while keeping the system size constant at $l=32$. We initialized the system with an aster in the polarization field, and varied the initial motor concentration $m_0$. Figure~\ref{fig:naster_bulk} shows the number of asters as a function of motor concentration. We find a critical motor concentration $m_0=1.5$ above which there are multiple asters emerging. 
Using the maximal wavevector determined Eq.\eqref{eq:simple_max_wavevector} by our linear stability analysis we can compute the motor concentration at which the corresponding wavelength $L_{\mathrm{max}}$ is half the system size, such that it can support two asters. 
The resulting critical motor concentration,
\begin{align}
   m_0= \frac{4 \pi ^2 (D_m+\kappa_p \Gamma_p)^2}{\zeta  (l/2)^2 v_0 },
    \label{eq:simple_max_c0_app}
\end{align}
has a value of $m_{0} = 1.3$, which agrees reasonably well with the critical motor concentration we find from our simulations.

% Temporarily moved to accommodate figure

\begin{figure}
    \centering
    \includegraphics[width=0.49\textwidth]{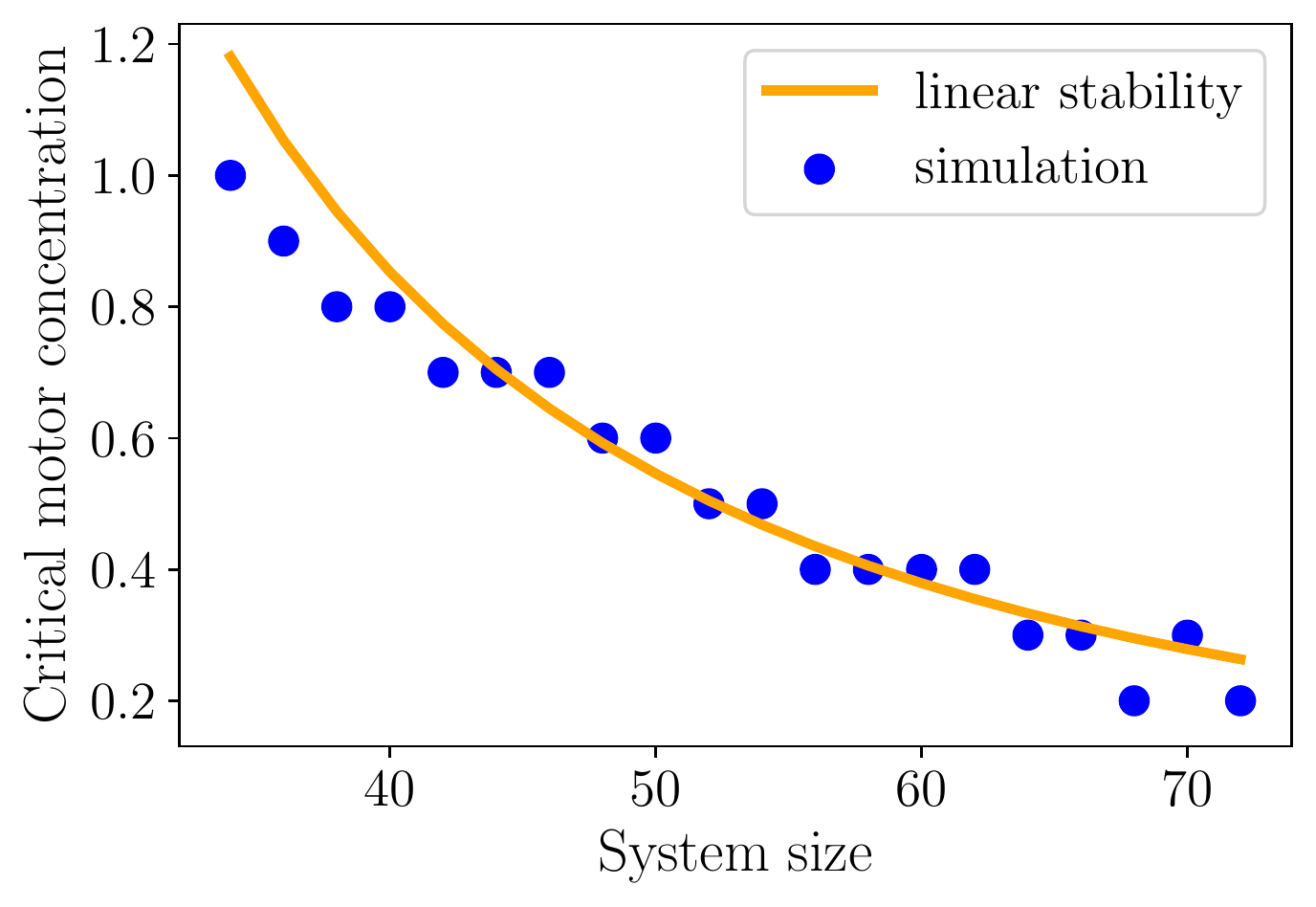}
    \caption{Critical motor concentration as a function of system size in a bulk simulation. Circles show the critical motor concentration from our simulations and the solid line show the result of our linear stability analysis.}
    \label{fig:critical_motor_systemsize}
\end{figure}

In Fig~\ref{fig:critical_motor_systemsize}, we show the critical motor concentration at which one aster becomes unstable as a function of systems size. We show critical concentrations from both our simulations and our linear stability analysis, which agree well. The critical motor concentration becomes smaller with larger system size.

\section{Steady state aster solution}
In order to find a steady state solution corresponding to an aster, we focus on the motor and polarization equations, which at steady state are given by, 
\begin{align}
    &0=-\zeta_{p} {\nabla}  m +\Gamma_{p} \kappa_p \Delta \bm{p},
    \label{eq:steadystate_p}
    \\
    &0=\nabla \cdot [ D_m \nabla m+v_{0}   (m \bm{p}) ].
    \label{eq:steadystate_m}
\end{align}
Here, we neglected binding kinetics in the motor equation as well as the decaying terms for the polarization $\bm{p}$ and set compressibility $\zeta_0 \rightarrow 0$. A solution to Equation~\eqref{eq:steadystate_m} can be obtained by satisfying, $\nabla \mathrm{ln}(m)=-v_{0} /D_m \bm{p}$. We insert this into the Laplacian of Eq.\eqref{eq:steadystate_p} and find a solution,
\begin{align}
    0=\lambda_m  m + \Delta \mathrm{ln}(m),
    \label{eq:Liouville_Bratu_Gelfand}
\end{align}
where $\lambda_m=\frac{\zeta_p v_0 }{\Gamma_p \kappa_p D_m}$.
This is known as the Liouville–Bratu–Gelfand equation\cite{liouville1853,bratu1914,gelfand1963} and has been solved for a number a physically relevant cases \cite{walker1915}.
A single aster solution is given by
\begin{align}
    &m= \frac{8}{\lambda_m  \left(|\bm{r}|^2+ \xi_{c}^{2}\right)^2},
    \label{eq:LBG_aster_m}
    \\
    &\bm{p}=  p_{0}\frac{4 \bm{r}}{|\bm{r}|^2+ \xi_{c}^{2}},
        \label{eq:LBG_aster_p}
\end{align}
with $\xi_{c}$ as the characteristic aster size and $p_{0} =\frac{D_m}{v_0}$.  % is this the solution found numerically?

This set of solutions allows for $m \bm{p}= \nabla \omega$ with
\begin{align}
    \omega= -\frac{8 p_{0}}{\lambda_m (\xi_{c}^{2} + \bm{r}^2)^2}.
        \label{eq:LBG_aster_p2}
\end{align}
We can then write the active term in the density kinetics of Eq.~\eqref{eq:psi_equation} as $-\zeta \nabla \cdot(m \bm{p}) =-\zeta \nabla^2 \omega $. The density kinetics are then expressible by the variational derivative of an effective free energy, $F_{\mathrm{eff}}= -(\zeta/\Gamma_{\psi}) \int d^{2}x \, \omega \psi + F_{\psi}$.

Using the expression  \eqref{eq:LBG_aster_p2} for $\omega$ in the free energy term, we can obtain the contribution of the $\zeta$-dependent active flux term to the droplet surface energy. By assuming that $F_{\psi}$ determines the circular droplet shape, such that $\psi({\mathbf x}) = 1$ for $|{\mathbf x}|< R$ and zero otherwise, we can calculate this contribution as,
\begin{align}
    \int d^{2}x \, \omega \psi = 2\pi \int_{0}^{R} dr \frac{8 p_{0}}{\lambda_m} \frac{r dr}{(\xi_{c}^{2} + \bm{r}^2)^2} = \frac{8 \pi p_{0}}{\lambda_{m} \xi_{c}^{2}} \frac{R^{2}}{(\xi_{c}^{2} + R^2)},
\end{align}
which in the limit of small droplet size compared to aster size, gives a total contribution to the effective free energy, $-(\zeta/\Gamma_{\psi}) \int d^{2}x \, \omega \psi
\simeq  -\zeta  \cdot \frac{8 \pi p_{0}}{ \Gamma_{\psi} \xi_{c}^{4}} R^{2}$. Including the equilibrium surface energy of a circular droplet, we get for the total effective free energy that determines droplet density, $(\gamma - \zeta  \cdot \frac{8 \pi p_{0}}{ \Gamma_{\psi} \xi_{c}^{4}}) R^{2}$, which leads to droplet instability when $\zeta$ is large enough.
%effective free energy for the active term in the density equation shown in the main text.

\end{document}